\documentclass[onecolumn]{article}
\usepackage{amsmath,amssymb}

\usepackage{hyperref}
\begin{document}
\title{A Dust Universe Solution to the Dark Energy Problem}
\author{James G. Gilson\quad  j.g.gilson@qmul.ac.uk\thanks{
 School of Mathematical Sciences,
Queen Mary University of London, Mile End Road, London E1 4NS,
United Kingdom.}}
%\date{January 27, 2006}
\date{December 19th  2005}
\maketitle
\begin{abstract}
Astronomical measurements of the Omegas for mass density, cosmological constant lambda and curvature k are shown to be sufficient to produce a unique and detailed cosmological model describing dark energy influences based on the Friedman equations. The equation of state Pressure turns out to be identically zero at all epochs as a result of the theory. The {\it partial\/} omega, $\omega_\Lambda$ for dark energy, has the exact value, minus unity, as a result of the theory and is in exact agreement with the astronomer's measured value. Thus this measurement is redundant as it does not contribute to the construction of the theory for this model. Rather, the value of $\omega_\Lambda$  is predicted from the theory. The model has the characteristic of changing from deceleration to acceleration at exactly half the epoch time at which the input measurements are taken. This is a mysterious feature of the model for which no explanation has so far been found. An attractive feature of the model is that the acceleration change time occurs at a red shift of approximately 0.8 as predicted by the dark energy workers. Using a new definition of dark energy density it is shown that the contribution of this density to the acceleration process is via a negative value for the gravitational constant, -G, exactly on a par with gravitational mass which occurs via the usual positive value for G.
This paper also contains an appendix on dark energy dynamics with its own abstract. 
\vspace{0.4cm}
\centerline{Keywords: Dust Universe, Dark Energy, Friedman Equations}
\centerline{PACS Nos.: 98.80.-k, 98.80.Es, 98.80.Jk, 98.80.Qc}
\end{abstract}
\section{Introduction}
\setcounter{equation}{0}
\label{sec-intr}
The work to be described in this paper is the generation and study of the theory that should go with the experimental work\cite{01:kmo,02:rie} of the astronomers who claim that their measurements indicate that the universe expansion is accelerating.  Assuming that this theory is based on the Friedman equations\cite{03:rind,43:nar}, I have found a solution to those equations that seems to be inevitable if the ideas put forward by the dark energy workers are to be realized as a consequence of the model. Very briefly their main conclusion from observational astronomy is that a universe expansion process is taking place now that changed from deceleration to acceleration at some time $t_c$ in the past. This change is identified roughly as occurring in association with events at a universe radius $r_c=r(t_c)$ with an observed red shift here and now in the range,  $0.5 <z<1$.  I shall use the subscript $c$ as indicating the time of {\it change\/} from deceleration to acceleration. They assume that the cosmological constant $\Lambda$ is positive and deduce from experiment that $k=0$.
\section{Accelerating Model}
\setcounter{equation}{0}
\label{sec-am}
Using Friedman's equations,
\begin{eqnarray}
8\pi G \rho r^2/3 & = & {\dot r}^2 +(k - \Lambda r^2/3) c^2\label{1}\\
-8 \pi GP r/c^2 & = & 2 \ddot r + {\dot r}^2/r +(k/r -\Lambda r) c^2 \label{2}
\end{eqnarray}
in the case $k=0$ and a positively valued $\Lambda$ we have,
\begin{eqnarray}
8\pi G \rho r^2/3 & = &  {\dot r}^2 - |\Lambda | r^2 c^2/3 \label{3}\\
-8\pi GP r/c^2 & = & 2 \ddot r + {\dot r}^2/r -|\Lambda | r c^2.\label{4}
\end{eqnarray}
I define a radial length $R_{\Lambda}$ associated with $\Lambda$ as
\begin{eqnarray}
R_{\Lambda} = |3/\Lambda|^{1/2} \label{5}
\end{eqnarray}
and use Rindler's constant $C$,
\begin{eqnarray}
C = 8\pi G \rho r^3/3\label{6}
\end{eqnarray}
to write the  first Friedman equation in the form,
\begin{eqnarray}
{\dot r}^2 = C/r + (rc/R_{\Lambda})^2.\label{7}
\end{eqnarray}
Having the quantity, Rindler's\cite{03:rind} constant $C$, in equation (\ref{7}) fixed at a constant value turns out to be an important asset when we come to integrate the Friedman equations. If as is usually taken to be the case, the mass density function $\rho$ is taken to be of uniform value throughout the universe at any given epoch, $t$, though having a value that depends on epoch through $r(t)$ as $\rho (r(t))$  and the quantity $4 \pi r^3/3 = V_U(r)$ is interpreted as the total 3-dimensional volume of the universe when its radius is $r$, then the mass of the universe will be given by
\begin{eqnarray}
M_U(r) = \rho (r) V_U(r) = 4 \pi r^3\rho (r)/3\label{7.1}
\end{eqnarray}
and then the constant $C$ takes the form
\begin{eqnarray}
C = 2 M_U(r)G.\label{7.2}
\end{eqnarray}
From equation (\ref{7.2}), it follows while $C$ can be kept constant for variations of the boundary radius $r$ of the universe we can incorporate an $r$ dependent gravitation {\it constant\/}, $G(r)$, provided an $r$ dependent mass, $M_U(r)$, is suitably chosen to conform with equation (\ref{7.2}). This point will be returned to when we discuss quantization of the integrated Friedman equations. In big bang models in which all the mass of the universe is created at the big bang event and thereafter keeps at a constant value $M_U$, $G$ must also keep at a constant value if the equation (\ref{7.2}) is to be retained for constant $C$. The equation (\ref{7.2}) can also be retained for constant $C$ in cases where the mass of the universe is generated over time provided  a suitable time dependent $G$ is incorporated to keep $C$ constant over time. Most of the work to follow applies to either of these cases so that we do not need to specify which form of $G$ is involved. However, we do need to distinguish between the two cases if we wish to examine the mass density function $\rho$ of the universe explicitly.
The accelerating universe astronomical observational workers\cite{01:kmo} give measured values of the three $\Omega s$, and $w_\Lambda$ to be
\begin{eqnarray}
\Omega_{M,0} &=&8\pi G\rho_0/(3 H_0^2)=0.25\label{9}\\
\Omega_{\Lambda,0} &=& \Lambda c^2/(3 H_0^2)=0.75\label{10}\\
\Omega_{k,0} &=& -kc^2/(r_0^2 H_0^2) =0,\ \Rightarrow k= 0,\label{11}\\
\omega_\Lambda &=& P_\Lambda/(c^2 \rho _\Lambda) = -1\pm\approx 0.3.\label{11.1}
\end{eqnarray}
Here the value of Hubble's constant will be taken to be
\begin{eqnarray}
H_0 =72\ Km\  s^{-1}\  Mpc^{-1}\label{12}
\end{eqnarray}
or in inverse seconds\footnote{Many decimal places will be used to keep track of minutely different values that can arise from different calculation routes}
\begin{eqnarray}
H_0 = 2.333419756287\times 10^{-18} \  s^{-1}.\label{13}
\end{eqnarray}
From these values it follows that Rindler's constant $C$ has the form and value
\begin{eqnarray}
C = \Omega_{M,0} H_0^2 r_0^3 = 1.361211939757935r_0^3\times 10^{-36}\label{14}
\end{eqnarray}
in terms of the radius now, $r_0$.
$\Lambda$ and $R_\Lambda$ will have the forms and values
\begin{eqnarray}
\Lambda =\Omega_{\Lambda,0} 3 (H_0/c)^2= 1.363097286965269 \times 10^{-52}\label{15}\\
R_\Lambda = |3/\Lambda |^{1/2} =1.483532963676604 \times 10^{26}.\label{16}
\end{eqnarray}
It is convenient to rearrange the first Friedman equation (\ref{3}) in the successive forms,
\begin{eqnarray}
8\pi G \rho r^2/3 + \Lambda r^2 c^2/3 =  {\dot r}^2\label{28}\\
8\pi G \rho r^2/3 + 8\pi G \rho_\Lambda r^2/3 =  {\dot r}^2\label{29}
\end{eqnarray}
and so identify a mass density $\rho_\Lambda $ which can be used to account for the cosmological constant contribution as an additional mass density along with the original $\rho$.  Thus
\begin{eqnarray}
8\pi G \rho_\Lambda = \Lambda c^2\label{30}
\end{eqnarray}
or
\begin{eqnarray}
\rho_\Lambda = \Lambda c^2/(8\pi G). \label{31}
\end{eqnarray}
Thus in these terms the first Friedman equation (\ref{3}) becomes
\begin{eqnarray}
8\pi G (\rho +\rho _\Lambda ) = 3(\dot r/r)^2 = 3 H(t)^2.\label{32}
\end{eqnarray}
The second Friedman equation (\ref{4}) can be written as
\begin{eqnarray}
8\pi Gr(-P/c^2 + \rho_\Lambda ) =2 \ddot r + {\dot r}^2/r.\label{33}
\end{eqnarray}
The introduction of the additional mass density $\rho_\Lambda$ as in equation (\ref{31}) to explain the mathematical appearance or existence of the cosmological constant $\Lambda$ as a physical contributor to the theory is the usual approach.  However, it is not necessarily the best way of physically accounting for the cosmological constant as the resulting equation (\ref{32}) has the built in implication of putting $\rho$ and $\rho _\Lambda $ on a par with respect to the kinematic quantity $(\dot r/r)^2 = H(t)^2$ and this may not be physically very relevant. An alternative approach with an alternative density, $\rho^\dagger _\Lambda$, will be discussed later.

It follows from the definition (\ref{31}) of $\rho_\Lambda$  that the mass density associated with $\Lambda$ has the same sign as does $\Lambda$ itself which for some theory constructs is taken as positive and in other theoretical constructs is taken as negative. The negative mass density case does present conceptual difficulties. Using an equation of state involving pressure is also conceptually difficult in this context because an equation of state of the form
\begin{eqnarray}
P_\Lambda = \omega_\Lambda c^2 \rho_\Lambda\label{34}
\end{eqnarray}
for negative $\Lambda$ and positive $\omega_\Lambda $ implies that the pressure $P_\Lambda$ is also negative. For positive $\Lambda $ and negative $ \omega_\Lambda $ it also implies that the pressure $P_\Lambda$ is negative and is the case that is used by the astronomers. This negative pressure is usually accommodated with some intellectual gymnastics. However, in the case of a partial pressure as in equation (\ref{34}) it is more easily acceptable\footnote{A confusion as between the total pressure $P$ and the partial pressure $P_\Lambda$ in the first version of this paper has been rectified in this present version.}. I show here that such conceptual difficulties are avoided in the cosmological model to be examined here. In the next section, I derive a solution to the Friedman equations that can accommodate all four of the dark energy researchers experimental values.
\section{Dust Model Solution}
\setcounter{equation}{0}
\label{sec-dms}
From equation (\ref{7}),
\begin{eqnarray}
(dt/dr)^2 &=& {    \frac{1}{C/r + (cr/R_\Lambda)^2}   }\label{61}\\
(dt/dr) &=& \pm {   \frac{1}{(C/r + (cr/R_\Lambda)^2)^{1/2} }   }\label{61.1}\\
t &=& \pm\int {   \frac{ dr }{ (C/r + (cr/R_\Lambda)^2)^{1/2}}  }.\label{62}\\
 &=& \pm\int {   \frac{ dr }{(  C/r )^{1/2}(1 + (c/R_\Lambda)^2r^3/C  )^{1/2}}}\label{63}\\
 &=& \pm\int { \frac{ r^{1/2}dr }{(C)^{1/2}(1 + (c/R_\Lambda)^2r^3/C)^{1/2}}}. \label{64}
\end{eqnarray}
The change of variable $r\rightarrow y$ with the standard integration form at equation (\ref{67}) gives equation (\ref{68}) as $t$ in terms of the transformed variable, $y$.
\begin{eqnarray}
r^3 &=& y^2 \label{64.1}\\
(2/3) d(r^{3/2})/dr &=& r^{1/2}\label{65}\\
r^{1/2}dr &=& (2/3)dy\label{66}\\
\int \frac{dy}{(1+ ay^2)^{1/2}} &=& a^{-1/2}\ln(a^{1/2}y + (ay^2 +1)^{1/2}),\label{67}\\
t &=& \pm (C)^{-1/2}(2/3)\int {\frac{dy}{  (1 + (c/R_\Lambda)^2y^{2}/C)^{1/2}} } \label{68}\\
a &=& (c/R_\Lambda)^2/C .\label{69}
\end{eqnarray}
Thus the following sequence of steps gives the integral evaluated in terms of  the transformed variable $y$, an inversion back to the original variable $r$ at (\ref{76}) and the introduction of the simplifying function $\theta _\pm(t)$ and the constant $b$ to arrive finally at the solution(\ref{77.2}).
\begin{eqnarray}
t &=& \pm (C)^{-1/2}(2/3) a^{-1/2}\ln(a^{1/2}y + (ay^2 +1)^{1/2})\label{69.1}\\
  &=& \pm (C)^{-1/2}(2/3) ((c/R_\Lambda)^2/C)^{-1/2}\nonumber\\
&\times &\ln(((c/R_\Lambda)^2/C)^{1/2}y + (((c/R_\Lambda)^2/C )y^2 +1)^{1/2})\label{69.2}\\
 \pm 3ct/(2R_\Lambda) & = & \ln(cy/R_\Lambda)C^{-1/2} + ((cy/R_\Lambda)^2/C +1)^{1/2}).\label{70}\\
 ((cy/R_\Lambda)^2/C + 1)^{1/2}& = &  \exp(\pm3ct/(2R_\Lambda))- (cy/R_\Lambda)C^{-1/2}\label{71}
\end{eqnarray}
\begin{eqnarray}
(cy/R_\Lambda)^2/C +1& = &  \exp(\pm3ct/(R_\Lambda))- 2\exp(\pm3ct/(2R_\Lambda)) (cy/R_\Lambda)C^{-1/2}\nonumber\\
 &+& (cy/R_\Lambda)^2C^{-1}.\label{72}\\
1 & = & \exp(\pm3ct/(R_\Lambda))\nonumber\\
&-& 2\exp(\pm3ct/(2R_\Lambda)) (cy/R_\Lambda)C^{-1/2}.\label{73}\\
cy/(R_\Lambda C^{1/2}) & = & \frac{   \exp( \pm 3ct/(2R_\Lambda))- \exp(\mp 3ct/(2R_\Lambda ))    }{2} \nonumber\\
&=&   \sinh (\pm3ct/(2R_\Lambda)  ).\label{74}\\
cy & = & R_\Lambda C^{1/2} \sinh (\pm 3ct/(2R_\Lambda)  )\label{75}\\
cr^{3/2} & = &  R_\Lambda C^{1/2} \sinh (\pm3ct/(2R_\Lambda)  )\label{76}\\
r(t) & = & (R_\Lambda /c)^{2/3} C^{1/3} \sinh^{2/3} ( \pm 3ct/(2R_\Lambda)). \label{77}\\
b & = &  (R_\Lambda /c)^{2/3} C^{1/3}\label{77.1}\\
\theta _\pm(t) &=& \pm 3ct/(2R_\Lambda) \label{77.3}\\
r(t) & = &  b\sinh^{2/3}(\theta_\pm (t) ).\label{77.2}
\end{eqnarray}
Equation (\ref{77}) or (\ref{77.2}) is a formula for the radius of an expanding {\it accelerating\/} universe conforming to all four measurements of the dark energy investigators.
\section{Characteristics of Model}
\setcounter{equation}{0}
\label{sec-com}
The velocity, $v(t)=\dot r(t) =dr(t)/dt$, of expansion as a function of $t$ is given by
\begin{eqnarray}
v(t)&=& (R_\Lambda /c)^{2/3} C^{1/3} (2/3)\sinh ^{-1/3}(\theta _\pm (t))\nonumber\\
&\ &\times \cosh (\theta _\pm(t))(\pm 3c/(2R_\Lambda))\label{78}\\
&=& \pm (Cc/R_\Lambda)^{1/3} \sinh ^{-1/3} (\theta _\pm (t)) \cosh ( \theta _\pm(t))\label{80}\\
&=& \pm (bc/R_\Lambda)\sinh ^{-1/3} (\theta _\pm (t))\cosh ( \theta _\pm(t)).\label{80.1}
\end{eqnarray}
The acceleration $ a(t) = \dot v(t) = \ddot r(t)$ is found to be
\begin{eqnarray}
a(t)&=& ((c/R_\Lambda )^{4/3} C^{1/3}/2)(3 \sinh ^{2/3} (\theta _\pm)\nonumber\\
&\ &- \cosh ^2 (\theta _\pm (t)) \sinh ^{-4/3}( \theta _\pm (t)))\label{80.12}\\
&=& b(c/(R_\Lambda) )^2\sinh ^{2/3} (\theta _\pm (t))(3 - \coth ^2 (\theta _\pm(t)))/2 .\label{80.2}
\end{eqnarray}
From equation (\ref{77}) and equation (\ref{80}), Hubble's constant is given as a function of $t$ as
\begin{eqnarray}
H(t)&=& \dot r /r =  \frac{ \pm (Cc/R_\Lambda)^{1/3} \sinh ^{-1/3}( \theta _\pm (t))\cosh (\theta _\pm(t))} { (R_\Lambda /c)^{2/3} C^{1/3} \sinh^{2/3}(\theta _\pm(t))} \label{81}\\
&=& (c/R_\Lambda ) \coth(\pm 3ct/(2R_\Lambda)).\label{81.1}
\end{eqnarray}
The measured value of $H(t)$ is given by equation (\ref{13}). Thus a consistent value for time now, $t_0$, is the time solution of (\ref{81.3}) at (\ref{81.4}) with a numerical value in seconds at (\ref{81.5})
\begin{eqnarray}
&\ &(c/R_\Lambda ) \coth(\pm 3ct/(2R_\Lambda))= 2.333419756287\times 10^{-18} \label{81.3}\\
t_0 &=&  (\pm 2R_\Lambda)/3c) \coth^{-1}((R_\Lambda/c )\times  2.333419756287\times 10^{-18})\label{81.4}\\
&=&4.34467334479058\times 10^{17}\quad s, \label{81.5}
\end{eqnarray}
using the value of $R_\Lambda$ from equation (\ref{16}).
From the formula for the acceleration (\ref{80.1}), it can be seen that there is a time $t_c$ at which the acceleration changes from negative to positive given by the time solution of (\ref{81.51}) at (\ref{81.52})
\begin{eqnarray}
\cosh ^2 (\theta(t)) \sinh ^{-4/3}(\theta(t)) &=& 3 \sinh ^{2/3} (\theta(t)) \label{81.51}\\
t_c= (2 R_\Lambda/(3 c)) \coth ^{-1}(3^{1/2})&=& 2.172336672394881 \times 10^{17}.\label{81.52}
\end{eqnarray}
From equations (\ref{86}) and (\ref{77}) we can find the value of $r_0 =r(t_0)$, the radius of the universe formally at time-now as
\begin{eqnarray}
r_0= (R_\Lambda /c)^{2/3} C^{1/3} \sinh^{2/3} ( \pm 3ct_0/(2R_\Lambda)). \label{81.6}
\end{eqnarray}
The evaluation of which requires that we have a value for Rindler's constant $C$. However, an inspection of the definition for this constant equation (\ref{6}) shows that to know this constant we require knowing $H_0$ for which I have assumed an input value at equation (\ref{13}) but it also requires knowing the present radius of the universe $r_0$ for which we have {\it not\/} got a measured value.
There is an alternative approach to finding the value for the present time, $t_0$, and radius, $r_0 =r(t_0)$, of the universe and thus getting the appropriate value for Rindler's constant.
From equations (\ref{6}) and (\ref{9}), we note that Rindler's constant can be expressed in the form
\begin{eqnarray}
C= \Omega_M H^{\dagger 2} r^{\dagger 3},\label{81.7}
\end{eqnarray}
where I am now using the dagger superscript to denote the present day radius of the universe and the present day value for Hubble's constant. The dagger notation is to emphasise the alternative route used to calculate  quantities such as $r^\dagger$.
From equation (\ref{77}) for $r(t)$ it follows that
\begin{eqnarray}
r^{\dagger 3} = (R_\Lambda /c)^2 C \sinh^2(\pm 3ct^\dagger/(2R_\Lambda)). \label{84}
\end{eqnarray}
The value for Rindler's constant, $C$, from equation (\ref{81.7}) can be substituted into equation (\ref{84}) to give
\begin{eqnarray}
1 = 0.25 (R_\Lambda/c)^2 H^{\dagger 2}\sinh^2 (3 c t^\dagger/(2 R_\Lambda)) \label{85}
\end{eqnarray}
having cancelled the $ r^{\dagger 3}= r^3 (t^\dagger)$ that occurs on both sides of the equation. Thus
\begin{eqnarray}
\pm 1 = 0.5 (R_\Lambda/c) H^\dagger \sinh (3 c t^\dagger/(2 R_\Lambda)). \label{86}
\end{eqnarray}
This can be solved for $t^\dagger$. Taking the plus sign with unity and using equation (\ref{81.1}) at time now as in (\ref{87}).
\begin{eqnarray}
\ H^\dagger = H(t^\dagger)= (c/R_\Lambda ) \coth(\pm 3ct^\dagger/(2R_\Lambda)). \label{87}
\end{eqnarray}
It follows that the value of $t^\dagger$ is given by the time solution of (\ref{87.1}) as $t^\dagger$ at (\ref{88}).
\begin{eqnarray}
2&=& \cosh (\theta (t^\dagger))\label{87.1}\\
t^\dagger &=& (2 R_\Lambda/3c) \cosh ^{-1}(2) = 4.344673344789258 \times 10^{17}\ s.\label{88}
\end{eqnarray}
Remarkably close to the value obtained for the original $t_0$ displayed again at equation (\ref{88.1}).
\begin{eqnarray}
t_0=4.34467334479058\times 10^{17}\quad s. \label{88.1}
\end{eqnarray}
However, the value of $t^\dagger$ was obtained without directly using the numerical value of $H$ at equation (\ref{81}) although it does indirectly depend on $H$ through the value of $R_\Lambda$ so that it can be used to check the  theoretical numerical value of Hubble's constant $H^\dagger = H(t^\dagger)$ using the theoretical formula for $H(t)$.
Thus the time $t^\dagger$ is greater than $t_c$, the acceleration change time, by an amount
\begin{eqnarray}
t^\dagger - t_c =2.172336672394881 \times 10^{17} \ s = 6.888434400035 \times 10^9\ y. \label{89}
\end{eqnarray}
The $t^\dagger$ notation from now on will be used for time-now rather than the original $t_0$ but this implies that the value of Hubble's constant should be given by formula (\ref{81.1}) as
\begin{eqnarray}
H(t^\dagger)= (c/R_\Lambda ) \coth(\pm 3ct^\dagger/(2R_\Lambda)) =2.333419756287 \times 10^{-18}\ s^{-1}. \label{89.2}
\end{eqnarray}
Comparing this with the experimental value $2.333419756287\times 10^{-18} \ s^{-1}$ at equation (\ref{13}), we see that the value coming from this theory via the indirect route through the value of $R_\Lambda$ is exactly the experimental value given at (\ref{13}). However, although we have found $t^\dagger$ at formula (\ref{89}), we still have not found the value of $r^\dagger = r(t^\dagger)$ that is necessary to find Rindler's constant $C$ and the constant $b$. It seems that mathematically $C$ is to be taken as arbitrary or alternatively $r^\dagger$ is to be taken as arbitrary or either $C$ or $r^\dagger$ is to be obtained from experiment or further some other theoretical consideration needs to be  used to obtain one or other of these two key constants. There are values ascribed to $r^\dagger$ from astronomical observation  based on a various speculative extrapolations and which are therefore not greatly reliable see Gravitation\cite{3:mis} page 738, box(27.4).  I have here decided to use a value for $r^\dagger$ that comes from an assumption about the dependence of the gravitation constant on $t^\dagger$ or on $ct^\dagger$. This assumption comes from a suggestion about a formula for  $G$ in terms of other physical constants of great numerical accuracy noticed by  {\it Ross McPherson\/}, see references \cite{33:mcp,32:gil}. His original suggestion  had dimensions different from $G$.  A generalised version of McPherson's suggestion but which has the usual dimensions associated with the gravitation constant $G$ is as follows
\begin{eqnarray}
G = \hbar ^2/(m_p^2 m_e c t_0), \label{89.21}
\end{eqnarray}
where $m_p$ and $m_e$ are the rest masses of the proton and electron respectively. A formula similar to (\ref{89.21}) for the gravitation constant was suggested by P. A. Dirac\cite{28:dir} in the thirties. There are a number of ways in which this formula can be interpreted. Taking $t_0$ to have one of values associated with the present age of the universe gives a very accurate value for $G$\cite{18:moh,39:gil,41:gil}. Otherwise $ct$ can be given a value associated with the present radius of the universe $r^\dagger$ multiplied by $\cos (\chi _G)$, the numerical constant, $\cos (\chi _G)\approx 1$ coming from gravitational coupling, which will also give a value for $G$. I have used this last option to develop a quantum theory for gravity\cite{42:gil,19:gil,28:dir} that successfully gives accurate formulae conformable with Dirac's large number hypothesis. This gives me some confidence in now inverting the formula (\ref{89.21}) and using the measured value for $G$ to supply the missing numerical value for $r^\dagger$ as
\begin{eqnarray}
r^\dagger = \hbar ^2/(m_p^2 m_e  G)=6.539532681821 \times 10^{25}, \label{89.22}
\end{eqnarray}
where the numerical constant, $\cos (\chi _G)\approx 1$, has been omitted as its values is very close to unity at nearly all times. The situation when it could make some significant difference, at the start of expansion of the universe when  $t\approx +0$, will be discussed in a future paper. The issues of reformulating quantum theory so that it is consistent with the theory of stochastic processes, the quantization of gravity theory and the consequent production of a quantized cosmology is discussed in references \cite{04:gil,05:gil,06:gil,07:edd,08:kil,09:bas,10:kil}. The use of the formula (\ref{89.21}) for $G$ in terms of quantum quantities in this work represents a weak partial quantization of the structure. A more complete quantization will be discussed in a later paper.
This can now be used to complete the numerical computation for the present theory by using the value for $r^\dagger$ in equation (\ref{89.22}) to give us Rindler's constant $C$ through equation (\ref{14}) which is repeated below with $r^\dagger$ replacing $r_0$ at (\ref{89.23}) and its numerical value is given at (\ref{89.4}). The value of the constant $b$ is then deduced at (\ref{89.5}).
\begin{eqnarray}
C &=& \Omega_{M,0} H_0^2 r^{\dagger 3} = 1.361211939757935r^{\dagger 3} \times 10^{-36} \label{89.23}\\
&=& 1.361211939757935r^{\dagger 3}\times 10^{-36}= 3.806871984611\times
10^{41} \label{89.4}\\
b &=& (R_\Lambda /c)^{2/3}C^{1/3}=4.534258713925 \times 10^{25}. \label{89.5}
\end{eqnarray}
I finish this discussion with a few remarks about the parametric values that arise in this theory. Firstly we can calculate the red shift, $z$, that would apply to light emitted at the acceleration change time $t_c$ from the universe boundary $r(t_c)$ to arrive at a present day observer as
\begin{eqnarray}
z= r(t^\dagger)/r(t_c)-1\approx 0.81712.\label{90}
\end{eqnarray}
This certainly strengthens the theory-observation cohesion in that it is a theory result within the limits suggested by the dark energy workers.
An interesting and {\it curious\/}  direct result from this structure is that the ratio of time now to the  time of acceleration change as calculated from their respective values is exactly $2$,
\begin{eqnarray}
t^\dagger/t_c =2.\label{90.1}
\end{eqnarray}
 One might have noticed the value of this ratio from equations (\ref{81.52}) and (\ref{88}), if one had at some time earlier come across the {\it exact\/} numerical relation (\ref{90.2}),
\begin{eqnarray}
\coth ^2(\theta (t_c)) &=& 3, \label{90.12}\\
\cosh (\theta (t^\dagger)) &=& 2,\label{90.13}\\
\sinh (\theta (t^\dagger)) &=& 3^{1/2},\label{90.13b}\\
H(t_c) &=& 3^{1/2}c/R_\Lambda\label{90.131}\\
H(t^\dagger) &=& c \coth (\theta(t^\dagger))/R_\Lambda\label{90.132}\\
\cosh ^{-1}(2) &=& 2  \coth ^{-1} (3^{1/2}).\label{90.2}\\
H(t) &=& H(t_c) \coth (\theta (t))/3^{1/2}\label{90.21}\\
H(t^\dagger) &=& 2 H(t_c)/3.\label{90.22}
\end{eqnarray}
The first equation above (\ref{90.12}) essentially defines the time $t_c$ of change of acceleration from  negative to positive and comes from equation (\ref{81.52}). The second equation above comes from equation (\ref{86}) or (\ref{87.1}) and defines the {\it time-now\/}, $t^\dagger$,  when the input measurements were made. The fourth equation (\ref{90.131}) comes from equations (\ref{87}) and (\ref{90.12}).
I have not been able to decide whether or not the result equation (\ref{90.1}) is some remarkable coincidence in values of the present day $t^\dagger $ and the past $t_c$ or some deep indication of a need to have more than just the past epoch value $t_c$ to be able to talk about a definite location in the time history of an event such as a universally distributed change from deceleration to acceleration. However, equation (\ref{90.21}) does clearly show up the background {\it mathematical\/} theory structure of the relation between $t_c$ and $t$ at equation (\ref{90.1}) in relation to the time evolution of the system and gives equation (\ref{90.22}) the Hubble equivalent of (\ref{90.1}) when $t=t^\dagger$ but, I have to admit, that I have not yet deciphered the physical significance of this mathematics as expressed by (\ref{90.1}) or (\ref{90.22}). It would not seem so remarkable if they, (\ref{90.1}) and (\ref{90.22}),  were just  approximate results but in the construction of this model they are numerically {\it exact\/} and unexpected.

The formula for the pressure from the Friedman equation (\ref{4}) is, using the definitions for $H$ and $R_\Lambda$ at equations  (\ref{81}) and (\ref{16}) in (\ref{92}),
\begin{eqnarray}
P&=&-(c^2 /(8 \pi G r))(2 \ddot r + {\dot r}^2/r -|\Lambda | r c^2)\label{91}\\
 &=&-(c^2/(8 \pi G))(2 \ddot r/r + {\dot r}^2/r^2 -|\Lambda | c^2)\label{91.1}\\
 &=&(-c^2/(8 \pi G))(2 \ddot r/r + H(t)^2 -3 (c/R_\Lambda)^2).\label{92}
\end{eqnarray}
From equations (\ref{77.2}), (\ref{80.1}) and (\ref{80.2}) we have
\begin{eqnarray}
2 \ddot r(t)/r(t)&=& (c/R_\Lambda)^2(3 - \coth ^2 ( \theta _+(t))) \label{93}\\
H(t)^2 &=&  (c/R_\Lambda)^2 \coth ^2 (\theta _+(t)) \label{94}\\
2 \ddot r(t)/r(t)+ H(t)^2 &=& 3(c/R_\Lambda)^2.  \label{94.1}
\end{eqnarray}
Substuting these formulae into the form for pressure equation (\ref{92}), we find complete cancellation on the right hand side to give an identically equivalent to zero total pressure at all times,
\begin{eqnarray}
P(t) &=&(3c^2/(8 \pi G))(- (c/R_\Lambda)^2+ (c/R_\Lambda)^2) \label{94.2}\\
P(t) &\equiv& 0 \Rightarrow \omega \equiv 0. \label{95}
\end{eqnarray}
Thus this system  rigorously describes a {\it dust\/} universe.
From equation (\ref{92}) or (\ref{94.2})  we see that the total pressure $P$ can be expressed as the sum of the two partial pressures $P_\Lambda$ and $P_M$, the first of negative sign and the second of positive sign, as
\begin{eqnarray}
P_\Lambda &= -&( 3c^2/(8 \pi G))(c/R_\Lambda)^2 <0 \label{95.6}\\
P_M &=&( 3c^2/(8 \pi G))(c/R_\Lambda)^2 >0 \label{95.7}\\
P &=& P_M + P_\Lambda. \label{95.8}
\end{eqnarray}
The choice of which pressure is associated with which part of the total pressure
being determined by the sign of that part so that normal gravitational mass gives negative acceleration or positive pressure, that is gravitational {\it attraction\/}. The $\omega_\Lambda$ that goes with the $\Lambda$ equation of state involves negative pressure or positive acceleration is thus given by
\begin{eqnarray}
\omega _\Lambda &=& P_\Lambda /(c^2 \rho _\Lambda)= -( 3c^2/(8 \pi G))(c/R_\Lambda)^2 /(c^2\rho _\Lambda) \label{95.9}\\
&=&  -1\label{95.10}
\end{eqnarray}
an exact value from theory for $\omega _\Lambda$ at the centre of the measurement range at equation (\ref{11.1}).
A simple interpretation of the force structure that is the cause of the acceleration (\ref{80.2}) that is operative in this model is obtained if we introduce a mass density $ \rho^\dagger$ to account for the cosmological constant $\Lambda$ as
\begin{eqnarray}
{\rho^\dagger}_\Lambda = \Lambda c^2/(4\pi
G)=2\rho_\Lambda.\label{96}
\end{eqnarray}
That is to say its values is twice the value of the usual density function at (\ref{31}).  The Friedman equation (\ref{4}) repeated below at equation (\ref{97}) with $P=0$, can then be expressed as equation (\ref{98}) through to equation (\ref{100}).
\begin{eqnarray}
0&=& 2 \ddot r + {\dot r}^2/r -\Lambda r c^2.\label{97}\\
\ddot r&=& \Lambda r c^2/2 -{\dot r}^2/(2r).\label{98}\\
& = & \Lambda r c^2/2 - (4 \pi G\rho r + \Lambda c^2 r/2)/3\label{98.1}\\
& = & (\Lambda c^2 - 4\pi G \rho)r/3\label{98.2}\\
& = &4\pi r G(\rho^\dagger_\Lambda -\rho)/3 \label{99.1}\\
& = &4\pi r^3 G(\rho^\dagger_\Lambda -\rho)/(3 r^2)\label{99}\\
& = &{M^\dagger}_\Lambda G/r^2 - M_UG/r^2\label{100}\\
M^\dagger_\Lambda & = & 4\pi r^3 \rho^\dagger_\Lambda/3\label{101}\\
M_U & = & 4\pi r^3 \rho/3\label{102}
\end{eqnarray}
Thus associating the density function $\rho^\dagger_\Lambda$ and consequent total mass $ M^\dagger_\Lambda $ within universe with the cosmological constant $\Lambda$ we get the very transparent formula for the dynamics under gravity, the {\it acceleration} that any particle would experience at the boundary of the universe being given by equation (\ref{100}). It tells us clearly that the normal mass density $\rho$ gravitationally causes the usual decelerating attraction of this particle to within the universe body whilst the dark energy mass density causes an acceleration that repels this particle towards the outside of the universe. The theoretical structure described here involves this as a naturally occurring effect which is built into this model. Consequently any need to introduce negative mass densities to describe the dark energy contribution together with the conceptually difficult concept of negative {\it total\/} pressure are removed from contention.
\section{Conclusions}
It has been shown that a subset of the measurements by the dark energy workers given at equations (\ref{9}), (\ref{10}) and (\ref{11}) together with an input value for Hubble's constant now  and the assumption that Rindler's constant, $C$, is an absolute constant lead to a unique solution of the Friedman equations. The fourth measurement from the dark energy observations at equation (\ref{11.1}) is not necessary for finding this solution. In fact, we have found that this last measurement with the partial $\omega _\Lambda $ value, $\omega_\Lambda =-1$,  is an exact result derivable from the unique model from the first three measurements. Thus the model derived  from general relativity via Friedman's equations is conformable to all four of the dark energy measurements (\ref{9}), (\ref{10}), (\ref{11}) and (\ref{11.1}).

\vskip 1.5cm
\large
\centerline{\Large {\bf Appendix}}
\centerline{\Large {\bf Dark Energy and its}}
\centerline{\Large {\bf Possible Existence in Particulate Form}}
\centerline{\Large {\bf in  a Friedman dust Universe}}
\centerline{\Large {\bf with Einstein's Lambda}}
\vskip0.2cm 
\centerline{1st September 2009}
\vskip 0.75cm  
\large \centerline{\bf Appendix Abstract}
\vskip 0.5cm  
It is shown that negatively gravitating {\it particles\/} can {\it consistently\/} be considered to exist and interact with normal positively gravitating particles in the contexts of general relativity and classical Newtonian gravitational theory. This issue arises from the discovery of {\it dark energy\/} which is considered to be causing an acceleration of the expansion of the universe. The issue is, can this dark {\it energy\/} occur in particulate form?  A related issue was studied in the fifties by Herman Bondi, (\cite{65:bon}) when he investigated the possible existence of {\it negative mass particles\/} in the general relativity context, long before dark energy appeared on the scene. He came to a paradoxical conclusion that seemed to rule out the actual physical existence of negatively gravitating particles. This paradox does not occur in this work because only positive mass particle are involved whatever their gravitational character may be. The structure of the differential equations that would apply in the case of a binary pair of opposite gravitational character components are used to show and explain how they can become consistent in general relativity or classical gravitation theory. This involves explaining a non-obvious relation between the principle of equivalence and Newton's {\it Action Equals Reaction\/} principle. A path structure for a mixed mass binary pair is set up which satisfies the equations of motion and does not have paradoxical properties. The force structure of the system is checked with a known classical dynamical test for the force per unit mass involved in the component particles  motions. This test is used to demonstrates that the basic assumptions of this theory are incorporated into the consequential orbital structure. An alternative to the {\it Action Equals Reaction\/} principle more appropriate to the astronomical situation is suggested.  An animation using Mathematica has been derived, and is available, and shows how a mixed gravity binary pair move under their mutual gravitational action.     
\vskip 0.5cm
\section{Appendix Introduction}
\setcounter{equation}{0}
\label{sec-aintr}
The work to be described in this paper is an application of the cosmological model introduced in the papers {\it A Dust Universe Solution to the Dark Energy Problem\/} \cite{45:gil}, {\it Existence of Negative Gravity Material. Identification of Dark Energy\/} \cite{46:gil} and {\it Thermodynamics of a Dust Universe\/} \cite{56:gil}. The negatively gravitationally characterised mass density involved from which such hypothetical particles might be formed has the value, $\rho^\dagger_\Lambda$, which is twice the Einstein dark energy density, $\Lambda c^2/(8\pi G)$. The issue to be examined in this paper is the form that the {\it dark energy negatively gravitating material\/} is likely to take. I shall assume that it is in the form of material particles like the normal positively gravitating particles and consider what consequences this has for the dynamics of this exotic material in relation to the normal gravitating material. The previous work on this topic does lead towards making this particulate assumption about the form of dark energy by showing that with the new definition, $\rho^\dagger_\Lambda$, for dark energy density it comes physically onto a par with positively gravitating material simply by its gravitating strength being denoted by -G times mass in place of +G time mass as is usually the case. 
All of this work and its applications has its origin in the studies of Einstein's general relativity in the Friedman equations context to be found in references (\cite{03:rind},\cite{43:nar},\cite{42:gil},\cite{41:gil},\cite{40:gil},\cite{39:gil},\cite{04:gil},\cite{45:gil}) and similarly motivated work in references (\cite{10:kil},\cite{09:bas},\cite{08:kil},\cite{07:edd},\cite{05:gil}) and 
(\cite{19:gil},\cite{28:dir},\cite{32:gil},\cite{33:mcp},\cite{07:edd},\cite{47:lem},\cite{44:berr}). The applications can be found in (\cite{45:gil},\cite{46:gil},\cite{56:gil},\cite{60:gil},\cite{58:gil}\cite{64:gil}). Other useful sources of information are (\cite{3:mis},\cite{44:berr},\cite{53:pap},\cite{49:man},\cite{52:ham},\cite{54:riz}) with the measurement essentials coming from references (\cite{01:kmo},\cite{02:rie},\cite{18:moh},\cite{61:free}).  Further references will be mentioned as necessary.  
\section{Positive, Negative Gravitating Particle Dynamics}
\setcounter{equation}{0}
\label{sec-pngp}
The term  {\it pure particle\/}  will be used with an extended sense of meaning either astronomical accumulations of mass with order of the amount of mass that might be found within a planet, a star or a galaxy or quantum elementary particle and accumulated structures formed  from elementary particles such as atoms or molecules. In other words, a pure particle can be mass-wise any physical object formed from a mass accumulation with some restrictions. I shall assume that particles can have the two possible states of either being gravitationally positive, $m^+$, and consequently causing  acceleration towards itself on all other particles in its vicinity or being gravitationally negative, $m^-$, and causing an acceleration away from itself on all other particles in its vicinity. All the particles involved are assumed to have {\it positive mass\/} and the restriction {\it pure\/} means that all such particles, whatever mass size will each have a definite gravitational coupling value, $\pm G$,  $G$ being Newton's gravitational constant. That is to say, for a massive particle of mass $m^\pm$ gravitating strength equals $\pm G m^\pm$. This is a provisional restriction that will certainly need to be lifted that I introduce to avoid mixed character particles that would involve changed values for $G$. Further restrictions on this very wide definition for possible masses will be discussed later. I shall designate the gravitationally characterised positive particles as being of red colour and gravitationally characterised negative particles as being of green colour for ease of reference and diagrammatic use and further represent the red particles with a $+$ superscript and the green particles with a $-$ superscript. I emphasise that both types of particle have {\it positive\/} rest mass. Thus a red particle induces, at distant points, accelerations {\it towards\/} itself in {\it all\/} other particles in its vicinity. A green particle induces, at distant points, accelerations {\it away\/} from itself in {\it all\/} other particles in its vicinity. This assumption about gravitational characterisation is very different from the electric characterisation for charged particles with charges $\pm e$ where oppositely characterised particles mutually attract and same characterised charges mutually repel, the distinction emphasised in the {\it all} for the gravitational case. This may seem to the reader to invite a paradox but it will be explained why this paradox does not arise.
The classical Newtonian equations that describes how a red particle behaves dynamically under the gravity field of another red particle are
\begin{eqnarray}
{\bf f}&=&m_1^+ d^2{\bf r}/dt^2=-Gm_1^+m_2^+ {\hat{\bf r}}/r^2\label{w14}\\
d^2{\bf r}/dt^2&=& ({\ddot r}-r{\dot\theta}^2){\hat{\bf r}}+(r{\ddot \theta} +2\dot r{\dot\theta})\hat{\bf t}\label{w15}\\
\alpha_{r,R}= {\hat{\bf r}\cdot d^2{\bf r}/dt^2} &=&({\ddot r}-r{\dot\theta}^2)= -Gm_2^+/r^2.\label{w16}\\
\alpha_{\theta,R} ={\hat{\bf t}\cdot d^2{\bf r}/dt^2} &=&(r{\ddot \theta} +2\dot r{\dot\theta})=r^{-1}(d(r^2\dot \theta)/dt)=0.\label{w17}
\end{eqnarray}
The force acting is the gravitational force originating from the positively gravitating particle $m_2^+$, regarded as fixed at its specific origin of coordinates, is directed towards it. That is to say, $m_2^+$ is at the tail of the vector ${\bf r}$ and $m_1^+$ is at the sharp end of ${\bf r}$ and usually in motion.
These equations contain much information about classical Newtonian gravity. Very striking is the fact that the two components of acceleration $\alpha_{r,R}$ and  $\alpha_{\theta,R}$ of the particle, $m_1^+$, do not involve the mass of this particle, in fact it could be red or green and in either case the equation would be unchanged. The mass of the red particle $m_2^+$ is involved in the $\hat{\bf r}$ component through the influence of the gravitational field. Thus the accelerating effect on any particle of any mass or any other characteristic under the influence of a red gravitating source is the same for all particles passing through the same position distant from $m_2^+$ at any time. Thus from classical dynamical theory, we obtain the essence of Einstein's {\it principle of equivalence\/}. The main conclusion, I wish to emphasise is that the acceleration, $\alpha_{r,R}$, is always negative with direction, ${-\hat{\bf r}}$ that is towards $m_2^+$, for any particle experiencing the effect of a positively characterised particle such as $m_2^+$ because $G$ and $m_2^+/r^2$ are both always positive. That is the induced acceleration is towards a positively gravitating source, $m_2^+$ whatever the gravitational character the subjected particle, $m_1^\pm$, may have, 
\begin{eqnarray}
\alpha_{r,R}=\frac{{\bf f}\cdot \hat{\bf r}  }{ m_1^+}&=&({\ddot r}-r{\dot\theta}^2)\le 0.\label{w18.1}
\end{eqnarray}
Let us now consider the case of a green particle fixed at the origin of coordinates being the source of the gravitational force on a red particle. For this case equation {\ref{w16}} has to be replaced with the negative $-G$ becoming $+G$  as in the next equation.      
\begin{eqnarray}
\alpha_{r,G}=\frac{{\bf f}\cdot {\hat{\bf r}}}{ m_1^+}=({\ddot r}-r{\dot\theta}^2)= +Gm_2^-/r^2.\label{w18}
\end{eqnarray}
This then implies also from equation (\ref{w16})
\begin{eqnarray}
\alpha_{r,G}=\frac{{\bf f}\cdot {\hat{\bf r}}}{ m_1^+}=({\ddot r}-r{\dot\theta}^2)\ge 0,\label{w19}
\end{eqnarray}
as all rest masses and $G/r^2$ are always positive. Thus a negatively gravitating  mass source particle induces acceleration away from itself for all particles in its vicinity regardless of any other characteristics they  may have. Particle, $m_2^-$, in equation {\ref{w18}} differs from the positively gravitationally characterised particle $m_2^+$ in equation {\ref{w16}} in that it is negatively gravitationally characterised.
It follows that for a binary system composed of a positive gravitating mass and a negative gravitating mass we need two equations of motion such as, 
\begin{eqnarray}
{\bf f}_1 &=&m_1^+ d^2{\bf r}_1 /dt^2=+Gm_1^+m_2^- {\hat{\bf r}_1}/r_1^2\label{w25}\\
d^2{\bf r}_1 /dt^2&=& ({\ddot r_1 }-r_1 {\dot\theta_1 }^2){\hat{\bf r}_1}+(r_1 {\ddot \theta_1 } +2\dot r_1 {\dot\theta_1 })\hat{\bf t}_1\label{w26}\\
\alpha_{r_1,G}&=&\frac{{\bf f}_1\cdot {\hat{\bf r}_1 }}{ m_1^+ }=({\ddot r_1 }-r_1 {\dot\theta_1 }^2)= +Gm_2^-/r_1^2\ge 0.\label{w27}\\
\alpha_{\theta_1,G}&=&\frac{{\bf f}_1\cdot {\hat{\bf t}_1}}{ m_1^+ }=(r_1 {\ddot \theta_1 } +2\dot r_1 {\dot\theta_1 })=0,\label{w28}
\end{eqnarray}
\begin{eqnarray}
{\bf f}_2&=&m_2^- d^2{\bf r}_2/dt^2=-Gm_1^+m_2^- {\hat{\bf r}_2}/r_2^2.\label{w291}\\  
d^2{\bf r}_2/dt^2&=& ({\ddot r_2}-r_2 {\dot\theta_2}^2){\hat{\bf r}_2}+(r_2{\ddot \theta_2} +2\dot r_2{\dot\theta_2})\hat{\bf t}_2\label{w301}\\
\alpha_{r_2,R}&=&\frac{{\bf f}_2\cdot {\hat{\bf r}_2}}{ m_2^- }=({\ddot r_2}-r_2{\dot\theta_2}^2)= -Gm_1^+/r_2^2\le 0.\label{w311}\\
\alpha_{\theta_2,R}&=&\frac{{\bf f}_2\cdot {\hat{\bf t}_2}}{ m_2^- }=(r_2{\ddot \theta_2} +2\dot r_2 {\dot\theta_2})=0,\label{w321}
\end{eqnarray}
The two sets of equations above refer to a system of two particles one, $m_1^+$, with a positive gravitating characteristic and one, $m_2^-$, with a negative gravitating characteristic, both have positive rest mass. In the first set $m_2^-$ is fixed at the origin of coordinates and is the source of the gravitational field that determines the dynamics of $m_1^+$ which is found at vector position ${\bf r_1}$ at time $t$ relative to the position of $m_2^+$.
In the second set $m_1^+$ is fixed at the origin of coordinates and is the source of the gravitational field that determines the dynamics of $m_2^-$ which is found at vector position ${\bf r_2}$ at time $t$ relative to the position of $m_1^+$. 
Thus there are two {\it space} frames of reference involved with their origins separated by the relative position vector, ${\bf r}= {\bf r}_1 -{\bf r}_2$, of the particles. The key quantities that determine the kinematic and dynamic behaviour of the total system are the four component accelerations  $ \alpha_{r_2,R}, \alpha_{r_1,R},\alpha_{r_2,G}, \alpha_{r_1,G}$ but being measured in different frames of reference it is difficult to see how their effects are to be combined. This is necessary if we are to understand how the particles interact.
However, Inspection of the two basic equations of motion (\ref{w25}) and (\ref{w291}), repeated below, which somehow or other are to hold  as a  simultaneous pair for the two particle system under examination,
\begin{eqnarray}
{\bf f}_1 &=&m_1^+ d^2{\bf r}_1 /dt^2=+Gm_1^+m_2^- {\hat{\bf r}_1}/r_1^2\label{w331}\\
{\bf f}_2&=&m_2^- d^2{\bf r}_2/dt^2=-Gm_1^+m_2^- {\hat{\bf r}_2}/r_2^2\label{w341}\\
{\hat{\bf r}_1}&=-&{\hat{\bf r}_2}\label{w341.1} 
\end{eqnarray}
the first impression is that they are incompatible as they seem to defy the Newtonian law that action and reaction should be equal. The first equation, gives the force from $m_2$ acting on $m_1$ . The second equation, gives the force from $m_1$ acting on $m_2$. Usually, if the action is taken to be ${\bf f}_1$, the reaction is regarded as being ${\bf f}_2$, then, $ {\bf f}_1 = -{\bf f}_2$, but here because of (\ref{w341.1}) it follows that $ {\bf f}_1 = {\bf f}_2$.  This is a well know paradox arising when trying to work with negatively gravitating particles. Subtle, you may say but nevertheless a source of great consternation in this area of work. It is clear from these equations that two negatively gravitating particles in interaction would also give the result $ {\bf f}_1 = -{\bf f}_2$, implying that their interaction satisfies Newton's third law so that we only have to consider the two situations of positively gravitating pairs and oppositely gravitating pairs to analyse and explain this situation. Thus we now have to consider the way this difficulty can be circumvented. We can get some direction in this investigation by considering how the similar well known problem in the study of the orbiting positive gravitating particles of a binary system in astronomy is handled.
There are also two equations of motion as in (\ref{w331},\ \ref{w341}),
 \begin{eqnarray}
{\bf f}_1 &=&m_1^+ d^2{\bf r}_1 /dt^2=-Gm_1^+m_2^+ {\hat{\bf r}_1}/r_1^2\label{x0}\\
{\bf f}_2&=&m_2^+ d^2{\bf r}_2/dt^2=-Gm_1^+m_2^+ {\hat{\bf r}_2}/r_2^2.\label{x1} 
\end{eqnarray}
Clearly there are no negatively gravitating particles involved here
and here we do have action and reaction equal because ${\bf {\hat r}}_1 =-{\bf {\hat r}}_2$ which implies ${\bf f}_1 = -{\bf f}_2$. The usual procedure is to get the two equations combined by defining the {\it centre of mass\/}, {\bf r}$_{cm}$, frame in place of the two obviously different frames used above as follows:
\begin{eqnarray}
{\bf r}_{cm}= \frac{m_1^+ {\bf r}_1 + m_2^+ {\bf r}_2}{m_1^++m_2^+}.\label{x2}
 \end{eqnarray}
In the absence of any {\it external\/} forces acting on a binary system, it is assumed that the center of mass can acquire no acceleration. Thus we can, by differentiating twice with respect to $t$ the centre of mass vector, deduce that
 \begin{eqnarray}
0=d^2{\bf r}_{cm}/dt^2\equiv m_1^+ d^2{\bf r}_1/dt^2 + m_2^+ d^2{\bf r}_2/dt^2,\label{x3}
 \end{eqnarray}
as the centre of mass denominator factors out.
It follows that
\begin{eqnarray}
0=d^2{\bf r}_{cm}/dt^2\equiv m_1^+ d^2{\bf r}_1/dt^2 + m_2^+ d^2{\bf r}_2/dt^2= {\bf f}_1 +{\bf f}_2 =0.\label{x4}
\end{eqnarray}
Thus the relations above become identities because the usual action equals reaction condition holds for the classical binary particle pair case. Clearly also, this condition as it stands does not hold for the mixed character particle pair case and it is fairly obvious why this is so. The centre of gravity has to replace the centre of mass in the mixed mass case. However, notably, in the classical situation the centre of mass and the centre of gravity of the systems coincide.
The centre of gravity in the mixed mass case can be defined as
\begin{eqnarray}
{\bf r}_{cg}&=& \frac{ G_+ m_1^+ {\bf r}_1 + G_- m_2^- {\bf r}_2}{G_+m_1^++ G_-m_2^-}=\frac{ m_1^+ {\bf r}_1 -  m_2^- {\bf r}_2}{m_1^+-m_2^-}\label{x5}\\
G_+&=&+G\label{x6}\\
G_-&=&-G.\label{x7}
\end{eqnarray}
In the mixed mass case, differentiating ${\bf r}_{cg}$ twice with respect to t  and assuming that the centre of gravity cannot acquire any acceleration in the absence of external forces gives
  \begin{eqnarray}
0=d^2{\bf r}_{cg}/dt^2\equiv m_1^+ d^2{\bf r}_1/dt^2 - m_2^- d^2{\bf r}_2/dt^2={\bf f}_1 -{\bf f}_2 =0, \label{x8}
\end{eqnarray}
because in this case, the condition, ${\bf f}_1 - {\bf f}_2=0$, contradicting action and reaction are equal holds. In the light of the form these structures take, it is perhaps tempting to conclude that all that need be done to get a consistent theory is to redefine the equality of action and reaction as the {\it magnitude\/} of action and reaction forces are always equal. I am reluctant to do this for reasons to be discussed later.
Using the basic two types of structure either centre of mass or centre of gravity orientated together with the separation vector of the two particles concerned ${\bf r}(t)={\bf r}_1(t)-{\bf r}_2(t)$ the position vectors for the two cases (\ref{x4}) and (\ref{x8}) can  respectively be each expressed in terms of ${\bf r}(t)$ in two equations,
\begin{eqnarray}
{\bf r}_1&=&{\bf r}_{cm}+\frac{m_2^+}{m_1^+ + m_2^+}{\bf r}(t)\label{x9}\\
{\bf r}_2&=&{\bf r}_{cm}-\frac{m_1^+}{m_1^+ + m_2^+}{\bf r}(t)\label{x10}\\
{\bf r}_1&=&{\bf r}_{cg}-\frac{m_2^-}{m_1^+ - m_2^-}{\bf r}(t)\label{x11}\\
{\bf r}_2&=&{\bf r}_{cg}-\frac{m_1^+}{m_1^+ - m_2^-}{\bf r}(t).\label{x12}
\end{eqnarray}
If we now differentiate these four equations through twice with respect to time assuming that the acceleration of the centre of mass or the centre of gravity is zero and then multiply through by the appropriate gravitationally subjected particle we get
\begin{eqnarray}
m_1^+d^2{\bf r}_1/dt^2 &=&+\frac{ m_1^+m_2^+}{m_1^+ + m_2^+} d^2{\bf r}(t)/dt^2= {\bf f}_1=-{\bf f}_2  \label{x13}\\
m_2^+d^2{\bf r}_2/dt^2 &=&-\frac{ m_2^+m_1^+}{m_1^+ + m_2^+} d^2{\bf r}(t) /dt^2={\bf f}_2  \label{x14}\\
m_1^+d^2{\bf r}_1/dt^2 &=&-\frac{ m_1^+m_2^-}{m_1^+ - m_2^-} d^2{\bf r}(t) /dt^2={\bf f}_1=+{\bf f}_2  \label{x15}\\
m_2^-d^2{\bf r}_2/dt^2 &=&-\frac{ m_2^-m_1^+}{m_1^+ - m_2^-} d^2{\bf r}(t) /dt^2={\bf f}_2.\label{x16}
\end{eqnarray}
The last four equations can be used to summarise results so far. 
The first two equations above are well known results from Newtonian gravitational theory and are used to find the astronomical orbits of binary star systems. The last two of the four are new results that describe the dynamics of binary mass systems composed of one positively gravitating mass together with one negatively gravitating mass and it has been shown that the new system seems to defy Newton's action and reaction law. This being the puzzling situation that the normally gravitating mass attracts the negatively gravitating mass whilst the negatively gravitating mass repels the normally gravitating mass. It has long been thought that this is a non reconcilable paradox that excludes the existence of negatively gravitating particles from the physical arena. An important point I wish to emphasise is that the {\it negatively gravitating\/} mass in this theory does not have negative mass. The gravitational negativity is an intrinsic property of its positive mass structure just as the negative charge of an electron is an intrinsic property of its positive mass structure. From the equations above the two component systems reduce to a single component equation in ${\bf r}(t)$ for the normal component pair and the mixed component pair. We note also that the usual {\it reduced\/} mass that arises in the binary system, $m_1G_+m_2/(G_+m_1+G_+m_2)$, in the mixed mass system is replace by $m_1G_-m_2/(G_+m_1+G_-m_2)$ and this will change sign if the mass character of the masses are interchanged. The form of the reduced mass for the mixed mass system also exposes a difficulty with the single equation of motion that arises for the mixed mass system.  If the two mass components $m_1^+$ and $m_2^-$ in such a system are equal their reduced mass becomes infinite. Thus rendering the reduced mass equation of state unusable. Consequently a binary mixed gravitating pair with {\it equal\/} masses must be excluded from discussion. I am sure that this difficulty has some fundamental significance but I do not know what it is. In the next section, I find a full solution for the paths of a mixed mass pair under their mutual gravitational interaction.
\section{Mixed Mass Pair Paths}
\setcounter{equation}{0}
\label{sec-mmpp}
In this section, I shall derive a complete solution to the dynamical problem of a positively gravitationally characterised mass and a negatively characterised mass moving in conjunction under their mutual gravitational interaction.  In the previous section, it was shown that such a system seems to violate the usual version of Newton's law of action and reaction being equal.
I shall work with a special case and show the solution does not involve bizarre features such as one particle chasing another to infinity which arose in the Bondi analysis. The production of this solution, I regard as something like a mathematics {\it existence\/} theorem, here showing that negatively gravitation particles can occur in nature in interaction with positively gravitating particles with unambiguous orbits, with a rational explanation and not contradicting general relativity or Newtonian dynamic. Repulsive inverse square law force is well known in the electromagnetic theory in the context of the  electron and positron interaction for example. The hyperbola is particularly interesting in the inverse square law context as it has two branches each with it own focus and with reference to one or other of its two foci, the two branches can represent an attracted particle path and a repelled particle path. In gravitation, theory the repelled particle path has hitherto been regarded as of no interest because gravity has always been thought to be only attractive. To give these remarks some mathematical basis let us first consider the pedal, $(r,p)$, equations of the two hyperbolic branches for a particle moving under inverse square law gravity,
\begin{eqnarray}
(b/p)^2 &=& 2 a /r +1\label{x17}\\
(b/p')^2 &=& 1-2 a /r' \label{x18}\\
1&=&(x/a)^2 - (y/b)^2.\label{x19}
\end{eqnarray}
The first equation above represents the orbit or branch of an hyperbola occupied by a particle being attracted to the focus within the orbit. This orbit is concave to the active focus. The second equation above represents the orbit or branch of the hyperbola occupied by a particle being repelled from a focus outside the orbit. This orbit is convex to the active focus. 
The second equation, (\ref{x18}) would usually be rejected in the gravitational context. The third equation involves both the branches and is represented in a frame of reference with the centre of the hyperbola at the centre of coordinates. Even though the receptor particle $ m_2^-$ is negatively gravitationally characterised, the numerical value of the attractive gravitational force from the particle $m_1^+$, fixed at the local focus, on a particle $m_2^-$ on the first branch above is given by,
\begin{eqnarray}
f_2=\frac{m_2^-m_1^+ G}{r^2} = \frac{m_2^-h^2}{p^3}dp/dr,\label{x20}
\end{eqnarray}
where $h=r^2 \dot\theta$ is the constant arrived at by integrating equation (\ref{w28}) or (\ref{w321}). Integrating equating (\ref{x20})with respect to $r$, we get
\begin{eqnarray}
\left(\frac{h}{p}\right)^2 = \frac{2 m_1^+G}{r}  + C.\label{x21}
\end{eqnarray}
If we multiply equation (\ref{x21}) through with $(b/h)^2$, we get
\begin{eqnarray}
\left(\frac{b}{p}\right)^2 = \left(\frac{b}{h}\right)^2 \frac{2 m_1^+G }{r} + C\left(\frac{b}{h}\right)^2
\label{x22}
\end{eqnarray}
and comparing (\ref{x22}) with (\ref{x17}), using the relation, $b^2=a^2(e^2-1)$, between b, a and e, the eccentricity, we find,
\begin{eqnarray}
C&=& \left(\frac{h}{b}\right)^2=\frac{h^2}{a^2(e^2 -1)}=\frac{(e^2 -1)(m_1^+G)^2}{h^2}= m_1^+G /a\label{x23}\\
a&=& \left(\frac{b}{h}\right)^2 m_1^+G = \frac{a^2(e^2 -1) m_1^+G }{h^2}\\ \label{x24}
a&=& \frac{h^2}{(e^2 -1)m_1^+G}.\label{x25}
\end{eqnarray}
Let us now consider the second branch of the hyperbola given by (\ref{x18}).
Firstly, suppose that we are not sure about the force that would have to be at the focus of the first branch to control the motion followed by a particle $m_1^+$ on the second path. Thus let us call this unknown or uncertain force force $G_?$. We now have to integrate the formula,
\begin{eqnarray}
f_1=\frac{ m_1^+ m_2^- G_?}{r^2} = \frac{m_1^+h^2}{p^3}dp/dr,\label{x26}
\end{eqnarray}
where $h=r^2 \dot\theta$ is the constant arrived at by integrating equation (\ref{w28}) or (\ref{w321}). Integrating the equation (\ref{x26}) with respect to $r$, we get
\begin{eqnarray}
\left(\frac{h}{p}\right)^2 = \frac{2 m_2^-G_?}{r}  + C.\label{x27}
\end{eqnarray}
If we multiply equation (\ref{x27}) through with $(b/h)^2$, we get
\begin{eqnarray}
\left(\frac{b}{p}\right)^2 = \left(\frac{b}{h}\right)^2 \frac{2 m_2^-G_? }{r} + C\left(\frac{b}{h}\right)^2
\label{x28}
\end{eqnarray}
and comparing (\ref{x28}) with (\ref{x18}), using the relation, $b^2=a^2(e^2-1)$, between b, a and e, the eccentricity, we find,
\begin{eqnarray}
C&=& \left(\frac{h}{b}\right)^2=\frac{h^2}{a^2(e^2 -1)}=\frac{(e^2 -1)(m_2^-G_?)^2}{h^2}= m_2^-G_? /a\label{x29}\\
a&=& -\left(\frac{b}{h}\right)^2 m_2^-G_? = -\frac{a^2(e^2 -1) m_2^-G_? }{h^2}\\ \label{x30}
a&=&- \frac{h^2}{(e^2 -1)m_2^-G_?}.\label{x31}
\end{eqnarray}
The two branches refer to the same hyperbola, (\ref{x29}), so that the value for $a$ obtained by the two different routes followed above should yield the same result. Thus from (\ref{x25}) and (\ref{x31}) it follows that
\begin{eqnarray}
\frac{a_{first\ route}}{a_{second\ route}}= 1 = 
-\frac{m_2^-G_?}{m_1^+ G_+}.\label{x32}
\end{eqnarray}
Here we seem to have a big problem, because we have been assuming that the force induced by a negative gravitating particle at a distant point involves $G_-=-G$ rather than $G_+ = G$ and here we see that substituting $G_-$ for $G_?$ gives the result
\begin{eqnarray}
\frac{a_{first\ route}}{a_{second\ route}}= 1 = 
\frac{m_2^-}{m_1^+ }.\label{x33}
\end{eqnarray}
This seems to be disastrous because it implies that the two masses have to be equal, a situation excluded earlier. However, it can also be taken to imply that we require two different sized hyperbolae each with its own $a$ value to get a consistent binary system. Thus we conclude that $G_?$ can be taken to equal to $G_-=-G$ provided we work with one hyperbola with a major axis $a_R$ and a second hyperbola with a major axis $a_G$ and such that
\begin{eqnarray}
\frac{a_R}{a_G} = 
\frac{m_2^-}{m_1^+}\ne 1. \label{x34}
\end{eqnarray}
However, the conclusion that two hyperbolae are necessary for the construction of a binary gravitational system is otherwise obvious. It would seem impossible to set up two particles moving on the branches of one definite hyperbola and at the same time have the particles fixed at the two available foci.

There are three {\it simple\/} ways to view a binary system. Two of them are from the rest mass frame of one or other of the pair and the third is from the centre of gravity frame for the pair. This observation is valid for all combinations of gravitationally characterised components. I have chosen to carry through the calculation here within the rest mass frame of the positively gravitating mass $m_1^+$. As earlier, I shall call $m_1^+$ the red component of the binary pair. In the rest frame of the red particle, the negatively gravitating green particle $m_2^-$ will be assumed to move on an hyperbola which as a whole is at rest, The path on which the green particle moves will be called the green path. This path will be an hyperbolic branch containing the stationary red particle fixed at the focus of that branch. The second fixed branch of this hyperbola does not play an active role so its focus will be displayed as an empty green circle in the diagrams $1$ on page $24$ and $3$ on pages $25$ to be found at \href{http://www.maths.qmul.ac.uk/~jgg/gil117.pdf}{QMUL Maths}. It will be helpful to refer to these diagrams for the following discussion.
Working in the rest frame of the red particle, which is by definition fixed in this frame, will have the consequence that the {\it path\/} of the red particle, the red path, will have to be in motion, rotating and translating while changing points on the red path remain attached to the definite fixed position of the red particle. Thus for the frame we are working in, the red path moves through the fixed red particle position rather than the red particle moves on the red path.
The green particle is on the green path but if it is to exert a repulsive force on the red particle it must also be at a remote focus of the red path. Thus mathematically we have to set up the two conditions:
\vskip 0.2 cm
\leftline{1 Red particle on local focus of green path}     
\leftline{2 Green particle on remote focus of red path}
\vskip 0.2 cm
The hyperbolic parameterisations for various hyperbolic branches with respect to the two foci posibilities are given below, 
\begin{eqnarray}
x_{LL}( \theta,a,e)&=&a (e-\cosh(\theta)) \label{x35}\\
x_{RL} (\theta,a,e)&=&-a (e+\cosh(\theta)) \label{x36}\\
x_{LR}(\theta,a,e)&=&a (e+\cosh(\theta)) \label{x37}
\end{eqnarray}
\begin{eqnarray}
x_{RR}(\theta,a,e)&=&-a (e-\cosh(\theta)) \label{x38}\\
y_{LL}(\theta,a,e)&=&a (e^2-1)^{1/2} \sinh(\theta) \label{x39}\\
&=&y_{LR} (\theta,a,e) \label{x40}\\
&=&y_{RL} (\theta,a,e ) \label{x41}\\
&=&y_{RR} (\theta,a,e )  . \label{x42}
\end{eqnarray}
The two letter subscripts $LL$ etc refer to the focus and branch involved respectively. For example $LR$ means left focus for origin of coordinates and right branch for path. This terminology is OK provided we rethink when a rotating hyperbola turns through more than $\pm \pi/2$. Thus to impose the conditions $(1)$ and $(2)$ into the mathematics of our binary system, we note that the separation distance $r_s$, between our red and our green particle can be expressed in two ways, in terms of the red particle parameters or in terms of the green particle parameters, each way giving the same numerical value for $r_s$ We are working in the red rest frame for the red particle and this is also the rest frame for the whole hyperbolic path of the green particle. This path will also have a fixed axis. However the red particle's hyperbolic path will be rotating in this frame so that we have to introduce its axial rotation away from the fixed axis of the green particle path. This rotation I shall call the angle $\beta$. $\beta$ is the angle between the green path's axis and the red path's axis. The positions of the various parameter components  are given by the blue lines in diagram $3$ on page $25$ to be found at \href{http://www.maths.qmul.ac.uk/~jgg/gil117.pdf}{QMUL Maths} together with identification positions $A$, $B$, $C$, $D$, $E$, $F$  defined below
\begin{eqnarray}
FA&=&x_{RL} \label{x42.1}\\
FD&=&y_{RL} \label{x42.2}\\
\angle BFA&=&\beta\label{x42.3}\\
\angle EFD&=&\beta\label{x42.4}\\
FB&=&x_{RL}\cos (\beta) \label{x42.5}\\
ED&=&y_{RL}\sin (\beta) \label{x42.6}\\
BA&=&x_{RL}\sin (\beta) \label{x42.8}\\
CB&=&y_{RL}\cos (\beta) \label{x42.9}\\
DC&=&x_{LL} \label{x42.91}\\
CA&=&y_{LL} \label{x42.92}
\end{eqnarray}
\begin{eqnarray}
FB-ED&=&DC\label{x42.93}\\
CB+BA&=&CA.\label{x42.94}
\end{eqnarray}
Thus using (\ref{x42.93}) and (\ref{x42.94}) and equating  components for
$r_s$ in the two parameterisations, we obtain
\begin{eqnarray}
x_{RL}( \theta_R,a_R,e_R)\sin(\beta)&+& y_{RL}(\theta_R,a_R,e_R)\cos(\beta)\nonumber\\
&=&\ \ \  y_{LL}(\theta_G,a_G,e_G) \label{x43}\\
-y_{RL}( \theta_R,a_R,e_R)\sin(\beta)&+&
x_{RL}(\theta_R,a_R,e_R)\cos(\beta)\nonumber\\
&=&\ \ \  x_{LL} (\theta_G,a_G,e_G) .\label{x44}
\end{eqnarray}
My objective is to produce a single nontrivial case of a mixed gravity binary system to establish that such systems can theoretically exist without internal contradiction or violation of physical laws. This can be most easily achieved by taking the simplest case. We have seen that the masses of the two components need be different and this translates into the $a's$ being not equal. The other parameter $e$ can be taken equal for the two orbits just to reduce notation and the arithmetic. Thus from now on I shall take $a_G = a$ and $a_R=0.3 a_G=0.3 a$. Then the basic equations become  
\begin{eqnarray}
x_{RL} ( \theta_R,0.3 a,e) \sin(\beta)&+& y_{RL}(\theta_R,0.3 a,e)\cos(\beta)\nonumber\\
&=&\ \ \  y_{LL}(\theta_G,a,e) \label{x45}\\
-y_{RL}( \theta_R,0.3 a,e)\sin(\beta)&+&
x_{RL}(\theta_R,0.3 a,e)\cos(\beta)\nonumber\\
&=&\ \ \  x_{LL} (\theta_G,a,e) . \label{x46}
\end{eqnarray}
These last two simultaneous equations can be solved for the pair of variables $\sin(\beta),\cos(\beta)$ to give
\begin{eqnarray}
\sin (\beta) &=&\frac{ x_{RL}( \theta_R,0.3 a,e)  y_{LL} (\theta_G,a,e)  - y_{RL}( \theta_R,0.3 a,e)  x_{LL}(\theta_G,a,e)  }{ x_{RL}( \theta_R,0.3 a,e) ^2+ y_{RL}( \theta_R,0.3 a,e)^2}\nonumber\\
 \label{x47}\\
\cos (\beta) &=& \frac{ y_{RL}( \theta_R,0.3 a,e)  y_{LL}(\theta_G,a,e)+ x_{RL}( \theta_R,0.3 a,e)  x_{LL}(\theta_G,a,e)  }{ x_{RL} ( \theta_R,0.3 a,e) ^2+ y_{RL}( \theta_R,0.3 a,e) ^2}.\nonumber \\
\label{x48}
\end{eqnarray}
The expanded versions for  $\sin (\beta) = S( \theta_R, \theta_G) $ and $\cos (\beta) =C( \theta_R, \theta_G)$  are given next in terms of the parameters, $\theta_R$ and $\theta_G$, for position on the two hyperbolae, 
\begin{eqnarray}
& & \sin(\beta) =\left(\frac{(e^2-1)^{1/2}}{0.3}\right) \left(\frac{\cosh(\theta_G)\sinh(\theta_R)-\sinh(\theta_G)\cosh(\theta_R)}{ (e\cosh(\theta_R)+1)^2}\right.- \quad\quad \nonumber\\
& & \quad\quad\quad\quad\quad\quad\quad\quad\quad\quad\quad\quad\quad\quad\quad\quad \left.\frac{ e(\sinh(\theta_G)+\sinh(\theta_R)) }{(e\cosh(\theta_R)+1)^2}\right)                   \nonumber\\
& & \quad\quad\quad= S( \theta_R, \theta_G).                           \label{x49}
\end{eqnarray}
\begin{eqnarray}
%& & \nonumber\\ 
& & \cos(\beta) =\frac{3}{10}\left(\frac{e^2(\sinh(\theta_R)\sinh(\theta_G)-1) -e(\cosh(\theta_R)-\cosh(\theta_G))}{ (e\cosh(\theta_R)+1)^2}\right. -\nonumber\\
& &\quad\quad\quad\quad\quad\quad\quad\quad\quad\quad\quad\quad\quad \left.\frac{ \sinh(\theta_R) \sinh(\theta_G)+ \cosh(\theta_R) \cosh(\theta_G)}{ (e\cosh(\theta_R)+1)^2}\right)                 \nonumber\\
& & \quad\quad\quad=C( \theta_R, \theta_G).                    \label{x50}  
\end{eqnarray}
Thus it is possible to find the relation between $\theta_R$ and $\theta_G$ for the binary system to hold together consistently by using the identity,
\begin{eqnarray}
\sin^2(\beta) +\cos^2(\beta)=1 \label{x51}  
\end{eqnarray}
and employing the functions above at (\ref{x49}) and (\ref{x50}). However, the formula that emerges by that route is intrinsic and complicated in the variables $ \theta_R$ and $ \theta_G$  and it is difficult to see how to extract an explicit relation giving for example a function of one in terms of the other such as, $\theta_R(\theta_G) $. A useful result that can easily be obtained from the $S$ and $C$ functions is the angle between the two orbit axes for any pair of positions. This is given by the inverse tangent 
\begin{eqnarray}
\beta (\theta_R, \theta_G)=\tan^{-1}\left(\frac{ S( \theta_R, \theta_G)}{ C( \theta_R, \theta_G)}\right) \label{x52}  
\end{eqnarray}
and can be used in finding the moving rotating path of the red particle in the green frame  or vice versa. If we return to the original pair of equations, we can find  the relation between the two position on curve parameters, $\theta_R$ and $\theta_G$, simply by squaring the two equations and adding the results together to obtain,
\begin{eqnarray}
r^2_{s,G}( \theta_R,e)&= &x_{RL}( \theta_R,a_R,e_R)^2+y_{RL}(\theta_R,a_R,e_R)^2\label{x53}\\
&=&x_{LL} (\theta_G,a_G,e_G)^2 + y_{LL}(\theta_G,a_G,e_G)^2 \label{x54}\\
&=& r^2_{s,R}( \theta_G,e ),\label{x55}
\end{eqnarray}
whilst eliminating $\beta $ at the same time. This result simultaneously defines the square of the separation distance, $ r^2_{s,G}( \theta_R)$, between the two particles, when viewed from the green particle or as, $r^2_{s,R}( \theta_G )$, when viewed from the red particle, $\theta_R $ varying from the green point of view and vice versa. Expanding the expressions in (\ref{x53}) and (\ref{x54}) we find,
\begin{eqnarray}
r^2_{s,G}( \theta_R,e)&= &(0.3a)^2 (e \cosh (\theta_R)+1)^2\label{x56}\\
r^2_{s,R}( \theta_G,e)&= & a^2 (e \cosh (\theta_G)-1)^2\label{x57}\\
r_{s,G}( \theta_R,e)&= & \pm 0.3 a (e \cosh (\theta_R)+1) \label{x56.1}\\
r_{s,R}( \theta_G,e)&= & \pm a (e \cosh (\theta_G)-1)\label{x57.1}\\
\theta_{R,\pm, \tilde\pm }(\theta_G) &=& \tilde\pm \cosh^{-1} (\pm \cosh(\theta_G)/0.3 \mp 1/(0.3 e) - 1/e) \label{x58.1}\\
\theta_{G,\pm,\tilde\pm}(\theta_R) &=&\tilde \pm \cosh^{-1} (\pm 0.3\cosh(\theta_R) + 1/e \pm 0.3/e).\label{x59.1}
\end{eqnarray}
The fifth equation above is the sought relation between $\theta_R $ and $\theta_G$ resulting from the equality of the two separation views from the first two equations.
The leading $ \tilde\pm $ at (\ref{x58.1})and (\ref{x59.1}) results from the evenness of the $\cosh$ function with the $\sim$ indicating that this $\pm$ is independent of the others,  so implying four possibilities. The sixth expression above is the inverse transformation of the fifth. The polar forms for the separation distances can be obtained by replacing  the parameters $\theta_R $ and $ \theta_G$ by functions of their angular equivalents $ \phi_R $ and $\phi_G$, say, 
\begin{eqnarray}
\phi_R(\theta_R) &=& \tan^{-1} \left(\frac{- y_{RL}(\theta_R,0.3 a,e)}{x_{RL}(\theta_R,0.3 a,e)} \right)\label{x59.2}\\
\phi_G(\theta_G )&=& \tan^{-1}\left( \frac{y_{RL}( \theta_G,a,e) }{ x_{RL}(\theta_G,a,e)} \right)\label{x60}\\
\phi(\theta_G )&=& \tan^{-1}\left( \frac{y_{LL}( \theta_G,a,e) }{ x_{LL}(\theta_G,a,e)} \right) .\label{x61}
\end{eqnarray}
Both expressions at (\ref{x56.1}) and (\ref{x57.1}) after the $\pm$ signs above are always positive because $e>1$ and $\cosh (x) >1$. Consequently, one might infer that the negative options can be ignored on the grounds of being unphysical, if the $r_{s,R/G}$ quantities represent physical distance. This turns out to be not the correct inference so that the final definitions of these quantities have to be such that they, being scalar distance, are carefully defined to come out positive. This will be explained. The inversions of the three equations above, the $\theta$s in terms of the $\phi$s are
\begin{eqnarray}
\theta_R (\phi_R) &=& \cosh^{-1} \left(\frac{e \tan^2(\phi_R) -\sec (\phi_R)(e^2 -1) }{e^2 - \sec^2 (\phi_R)} \right)\label{x62}\\
\theta_G(\phi_G )&=& \cosh^{-1} \left(\frac{e \tan^2(\phi_G) +\sec (\phi_G)(e^2 -1) }{e^2 - \sec^2 (\phi_G)} \right)\label{x63}\\
\theta(\phi)&=& \cosh^{-1} \left(\frac{e \tan^2(\phi) -\sec (\phi)(e^2 -1) }{e^2 - \sec^2 (\phi)} \right)) .\label{x64}
\end{eqnarray}
As a result of these equations, we can re-express the formulae for the distance between red and green particles, the  $r_{s,R}(\theta_G) $ and $r_{s,G}(\theta_R)$ functions at (\ref{x56.1}) and (\ref{x57.1}), in terms of the angles $\phi_G$ and $\phi_R$ respectively.
\begin{eqnarray}
r^\prime_{s,G}( \phi_R,e) & =& \pm 0.3 a (e \cosh (\theta_R(\phi_R))+1)\label{x65}\\
&\rightarrow & \left| 0.3 a\left(   \left(\frac{e^2 \tan^2(\phi_R) -\sec (\phi_R)e(e^2 -1) }{e^2 - \sec^2 (\phi_R)} \right)+1\right)\right| \label{x66}\\
&=&\left| \frac{0.3a (e^2-1)}{\cos (\phi_R)e +1}  \right|\label{x66.1}\\
 r^\prime_{s,R} ( \phi_G,e) & =& \pm a(e \cosh (\theta_G)-1)\label{x67}\\
&\rightarrow &\left| a\left(\left( \frac{e^2 \tan^2(\phi_G) +\sec (\phi_G)e(e^2 -1) }{e^2 - \sec^2 (\phi_G)} \right) -1\right)\right|\label{x68}\\
&=&\left| \frac{a(e^2-1)}{\cos (\phi_G)e -1}  \right|\label{x68.1}
\end{eqnarray}
the primes being introduced to differentiate these new functions of $\phi$ from the original functions of $\theta$. The $\pm$ options have been assigned so that both these length functions are positive as indicated by the modulus sign, $||$ enclosing their final definitions. This is the point I suggested earlier should be treated with care as the introduction of the $\phi$ versions with the inverse $\cosh$es seems to open up the possibility for negative values. The relations between the $\phi_R$s and $\phi_G$s corresponding to the relations between the $\theta_R$s and $\theta_G$s at (\ref{x58.1}) and (\ref{x59.1}) which make the two lengths equal are obtained easily by taking the two lengths {\it as equal\/} and then solving for either $\phi_G$ or $\phi_R$ and are
\begin{eqnarray}
\phi_G (\phi_R) &=& \cos^{-1} ( (\cos (\phi_R)/0.3 +1.3/(0.3 e))) \label{x69}\\
\phi_R (\phi_G) &=& \cos^{-1} ( (0.3\cos (\phi_G) -1.3/e)) \label{x70}
\end{eqnarray}
This completes most of the technicalities and leaves us with the two distances, $r^\prime_{s,G}(\phi_R)$ and $r^\prime_{s,R}(\phi_G)$, between the red and green particles as would be seen by an observer on the green particle in terms of the red particle's angular parameter $\phi_R$ and an observer on the red particle in terms of the green particle's angular parameter $\phi_G$ respectively. The green observer sees the red particle and its path and the red observer sees the green particle and its path. However, these distances are not of equal magnitude unless the basic $\theta_R$ and the basic $\theta_G$ are related by the formula (\ref{x58.1}) or the formula (\ref{x59.1}) or the equivalent $\phi_R$ and $\phi_G$  are equivalently related.
\section{Assembling the Paths}
\setcounter{equation}{0}
\label{sec-atp}
Most of the mathematics for this problem has now been completed. We have seen that in the rest frame of the red particle, the path of the green particle is also at rest, the red particle being at the focus of this hyperbolic path. It is also established that the axis of the path of the red particle will have to have rotated relative to the green path axis by the angle $\beta$ given at (\ref{x52}). Concentrating on the view of the situation from the rest frame of the red particle in which the path of the green particle is a simple hyperbolic branch also at rest, it is convenient to introduce a rotated frame, relative to this at the angle $\beta$ to it, in which the green particle is at rest and the path of the red particle is also at rest. To do this, I introduce the rotation-translation transformation of coordinates based at the empty green focus  which is at (0,0),
\begin{eqnarray}
X_{rot}(x,y,d_1,\beta)&=&(x\cos (\beta) -y\sin (\beta)) +d_1\label{x71}\\
Y_{rot}(x,y,d_2,\beta)&=&(x\sin (\beta) +y\cos (\beta)) +d_2\label{x72}\\
d_1&=&x_{RL}(\theta_G, a,e)\label{x73}\\
d_2&=&y_{RL}(\theta_G, a,e),\label{x74}
\end{eqnarray}
where $d_1$ and $d_1$  are displacements following the rotation that take the empty green focus to the momentary position of the green particle assumed to be at parameter value $\theta_G$. Thus this transformation takes us into a reference frame in which the green particle is at the origin of the transformed x-axis which is at the angle $\beta$ relative to the axis of the green parabolic path. In other words, the x-axis of this frame is on the x-axis of the red hyperbola with the green particle on its active focus. Thus to find the parametric equation for the red particle path all we have to do is place its parametric representation into the rotations  $(x,y)$ coordinate position and replace the $rot$ subscripts with $red$ to indicate we now have the two parametric components for the path of one of the branches of the moving red hyperbola as functions of $\phi_R$ and $\theta_G$.
\begin{eqnarray}
X_{red}( \phi_R, \theta_G)&=&  x_{RL}(\theta_R (\phi_R),0.3a,e)\cos (\beta (\theta_R (\phi_R), \theta_G))
\nonumber\\
&-& y_{RL}(\theta_R ( \phi_R),0.3a,e)\sin (\beta (\theta_R (\phi_R), \theta_G)))\nonumber\\
&+& x_{RL}( \theta_G, a,e)\label{x75}\\
Y_{red}(\phi_R, \theta_G)&=&x_{RL}( \theta_R (\phi_R),0.3a,e)\sin (\beta (\theta_R (\phi_R), \theta_G)) \nonumber\\
&+& y_{RL}(\theta_R (\phi_R),0.3a,e)\cos (\beta (\theta_R (\phi_R), \theta_G))) \nonumber\\
&+& y_{RL}(\theta_G, a,e),\label{x76}
\end{eqnarray}
where $\theta_R$ has been related to its angular equivalent $\phi_R$ and $\beta$ has been replaced by the function of $\theta_R$ and $\theta_G $ at (\ref{x52}) repeated below.
 \begin{eqnarray}
\beta ( \theta_R, \theta_G)=\tan^{-1}\left(\frac{ S( \theta_R, \theta_G)}{ C( \theta_R, \theta_G)}\right). \label{x77}  
 \end{eqnarray}
The other co-moving branch is given by
\begin{eqnarray}
X_{red}'( \phi_R, \theta_G)&=&  x_{RR}(\theta_R (\phi_R),0.3a,e)\cos (\beta (\theta_R (\phi_R), \theta_G))
\nonumber\\
&-& y_{RR}(\theta_R ( \phi_R),0.3a,e)\sin (\beta (\theta_R (\phi_R), \theta_G)))\nonumber\\
&+& x_{RL}( \theta_G, a,e)\label{x78}\\
Y_{red}'(\phi_R, \theta_G)&=&x_{RR}( \theta_R (\phi_R),0.3a,e)\sin (\beta (\theta_R (\phi_R), \theta_G)) \nonumber\\
&+& y_{RR}(\theta_R (\phi_R),0.3a,e)\cos (\beta (\theta_R (\phi_R), \theta_G))) \nonumber\\
&+& y_{RL}(\theta_G, a,e),\label{x79}
\end{eqnarray}
with the same adaptations as for the first branch.
With the parameterisation of the moving hyperbolic path for the red particle obtained at (\ref{x75}) and (\ref{x76}), it is possible to plot that path which has the moving green particle at its active focus and so give a clear diagrammatic picture for any chosen position, the value of $\theta_G$, of the green particle on it own path which you will recall is fixed, together with the red particle at its left focus, in the basic reference frame on which we are concentrating. The fixed path branches for the green particle is given by the parameterisations,
\begin{eqnarray}
X_{green}(\phi_G) &=&x_{RL}(\theta_G (\phi_G) ,a,e)\label{x80}\\
Y_{green}(\phi_G) &=&y_{RL}(\theta_G (\phi_G),a,e)\label{x81}\\
X_{green}'(\phi_G) &=& x_{RR}(\theta_G (\phi_G),a,e)\label{x82}\\
Y_{green}'(\phi_G) &=&y_{RR}(\theta_G (\phi_G),a,e).\label{x83}
\end{eqnarray}
We now have all the basic mathematical structure to plot and analyse the path structure and the relative motions of the particles in terms of the angular changes of $\phi_R$ and or $\phi _G$. Most of the emphasis here has been the view of the binary pair of an observer fixed to the rest frame of the red particle and the accompanying fixed hyperbolic path of the green particle. From that frame of reference the observes sees only the motion of the green particle around its orbit. It is not difficult to recast the whole structure in terms of the view of an observer on the green particle and his accompanying fixed hyperbolic orbit of the red particle. Further these two basic views can then be put together using the definitions of the centre of mass or centre of gravity systems to give an observers  view from either the centre of mass or the centre of gravity frame. However, only the last of these reference frame possibilities could be taken to be truly inertial. Using the {\it Mathematica\/} animation process, I have produced a simulation of the movement of the mixed character binary pair with $301$ frames. One of these frames has been used to produce the diagram on page $24$. The full simulation in mathematica note book language can be downloaded in the file $mixmass.nb$ from my website at, \href{http://www.maths.qmul.ac.uk/~jgg/mixmass.nb}{QMUL Maths}. It can now be shown that the structure of the force configurations for this binary pair composed of oppositely gravitationally characterised particles does conform to the correct conditions. They are that the red particle exerts a gravitational  attraction on the green particle and the green particle exerts a gravitational repulsion on the red particle. This can be done by using a well know result from classical dynamics and inverse square law forces which recovers the force involved from just knowing the distance in detail of a possibly influenced particle  from the force source as a function of the angle $\phi$. The formula required is derivable from (\ref{w16}) and is
\begin{eqnarray}
\alpha_{r,R}&=& {\hat{\bf r}\cdot d^2{\bf r}/dt^2}  = ({\ddot r}-r{\dot\phi}^2) = -Gm_2^+/r^2 (t)\nonumber\\
&=& h^2 u^2(\phi)( d^2 u(\phi)/d\phi^2 + u(\phi))= -Gm_2^+u^2(\phi) \label{x84}
\end{eqnarray}
\begin{eqnarray}
u(\phi)&=& 1/r(t) \label{x85}\\ 
h&=& r^2(t){\dot\phi}^2, \label{x86}
\end{eqnarray}
where normally $m_2^+$, as indicated, the gravitating source would be positively gravitating. $h$ is the constant arising from integrating the transverse component of acceleration $\alpha_{\phi,R}$ on the assumption that the mutual interaction of the particle pair only acts along their line of separation. As we are working with the inverse square law of gravitation $u^2(\phi)$ appears twice in the formula (\ref{x84}) and so can be cancelled from both sides of the second equality to define $ G_{type} (\phi)$ and to give,
\begin{eqnarray}
G_{type} (\phi)=h^2( d^2 u(\phi)/d\phi^2 + u(\phi))= -Gm_2^+,\label{x87} 
\end{eqnarray}
where $u(\phi)$ is the inverse of the magnitude of a position vector for distant points originating at the position of the mass $m_2^+$.  This shows immediately, because of the $-G$, that we are dealing with an attraction towards the source under gravity which in the past has been thought to always hold. The detailed steps in deriving the formula (\ref{x87}) outlined above can be found in A. S. Ramsey's book (\cite{66:ram}). They are elementary but I think this formula needs to be treated with care because it is rather profound. In particular, it should be noted that the force involved on the right hand side above is the force originating from the particle $m_2$ which is at the origin of coordinate from which the radial length, $r(\phi)=1/u(\phi)$, is measured and which is also the magnitude of a vector,  ${\bf r}(\phi )$, which has its tail at the same origin. However, the angular parameter $\phi$ on which it depends refers to the angular component of acceleration at the sharp end of the same vector and is involved with the path where the induced distant acceleration field may be influencing other particles but even so this $\phi$ is an angular parameter measuring an angle as perceived by an observer at the source of the force.  According  to general relativity all such other particles at the sharp end of the vector ${\bf r}(\phi)$ will experience the same acceleration due to the gravitational field of the particle $m_2$. So there is some sense that general relativity is at least a part feature of the formula. Thus when using this formula to find the nature of the force involved we must not lose sight that it refers exclusively to the source location of the force. The formula determines how the source observer sees the process at a distance but it contains no feedback from the distant events. This has the consequence that the function $u(\phi)$ which is used in the formula should be specific to the source as origin which one is trying to interpret. That is to say, it should not be in a form which makes it equal to the distance between the binary pair as observed from the other member of the pair. If the physically equal distances are made mathematically equal before using the formula, The two $u(\phi)$ functions become {\it identically\/} equal and so the differential form in the expression (\ref{x87}) gives the same result for the two binary members. The significance of this formula is that it gives information about the local source force, whether attractive or repulsive, valid for all receptor particles at whatever distance from the source they may be and whatever gravitational character they may have. This independence on receptor distance is indicated in the diagrams $1$ and $2$ on page $25$ to be found at \href{http://www.maths.qmul.ac.uk/~jgg/gil117.pdf}{QMUL Maths} by plotting the formula value with corresponding value of $u(\phi)$. Thus we can use the raw functions of $\phi_R$ and $\phi_G$ at (\ref{x66.1}) and (\ref{x68.1}) to determine the type of gravity originating from the green particle or the red particle respectively. These expressions are repeated below   
  \begin{eqnarray}
r^\prime_{s,G}( \phi_R,e) & =& \left| \frac{0.3 a(e^2-1)}{\cos (\phi_R)e +1}\right| =1/u_{Green}(\phi_R,e)\label{x88}\\
 r^\prime_{s,R} ( \phi_G,e) & =& \left| \frac{a(e^2-1)}{\cos (\phi_G)e -1}\right|=1/u_{Red}(\phi_G,e),\label{x89}
\end{eqnarray}
with the second equalities giving the appropriate $u (\phi,e)$ function.
If we evaluate the first and second derivatives of $u_{Green}(\phi_R,e)$ and $ u_{Red}(\phi_G,e)$, with respect to $\phi_R$ and $\phi_G$ respectively and then use them to find the value given by $G_{type}$ for the two cases, we find 
\begin{eqnarray}
G_{type} (\phi_R)&=&h^2( d^2 u_{Green}(\phi_R,e)/d\phi_R^2 + u_{Green}(\phi_R,e))>0\label{x90}\\
G_{type} (\phi_G)&=&h^2( d^2 u_{Red}(\phi_G,e)/d\phi_G^2 + u_{Red}(\phi_G,e))<0,\label{x91} 
\end{eqnarray}
implying that $ G_{type} (\phi_R)= Gm^+$ and $ G_{type} (\phi_G)= -Gm^-$ and thus  showing, as expected,  that the red particle will attract all other particles in its neighbourhood and the green particle will repel all other particles in its neighbourhood. The situation can be clearly seen from plots of $ G_{type} (\phi_R)$ for the two cases that are given at diagrams $1$ and $2$ on page $25$ to be found at \href{http://www.maths.qmul.ac.uk/~jgg/gil117.pdf}{QMUL Maths}. The corresponding $u (\phi)$ for the two cases is plotted with the $G_{type}(\phi)$ block function and this shows clearly how the character of any source does not depend on $r (\phi)=1/u(\phi)$, the distance of any other particle in its vicinity whatever character that particle may have.
\section{Conclusions}
\setcounter{equation}{0}
\label{sec-con}
The work in this paper makes strong {\it theoretical\/} evidence for the possible existence of {\it particulate negative gravitating\/} positive mass particles. This has been achieved by showing that negatively gravitating particles can be involved in composite systems and follow orbits in conjunction with positive gravitating particles in much the same way as the later type of usual particles can form systems and gravitationally interact. In the specific case examined here, this new orbital system type does  not have the bizarre features that were noted in the fifties studies by Herman Bondi. In the light of this special but typical case, it seems reasonable to assume that if such particles do exist their interaction generally with the normal type of particles would be theoretically predictable and not remarkably odd. Of course, I have not given any physical evidence that such particle do exist.

There are a number of interesting issues that arise from this work.  One such issue is the meaning and significance of the term {\it particle\/}. We talk about particle in quantum theory and at the other end of the scale, in cosmology, galaxies are often regarded as particles. In the definition of particle that has been used in this work the whole of this vast range has been included because it is generally believed that all such {\it particles\/} whether they have rest mass or not are influenced by and can cause gravity fields. However we think we know that negative gravity particles, if they exist at all, will be very rare entities. It is a fact that antiparticles in general are very rare, a fact that is regarded as a great mystery. The question is, where are the missing antiparticle? Some scientists believe that normal and anti particle were created together in equal numbers at an early stage of the evolution of the universe but now most of the antiparticle seem to have disappeared. This prompts me to make the speculation that {\it dark energy\/} is composed of those missing antiparticles and further those antiparticles that we do detect are manifestations of the local sea of dark energy particles and like them are also anti gravitational. This view is also supported by the accepted cosmological fact that at some time $t_c$ in the evolution of the universe there was equal quantities of normal mass and dark energy mass within the boundaries of the universe. This time could be the moment of creation of the equal anti and normal particle densities. The same identificaton for dark energy was put forward as a conjecture by {\it D. S. Hajdukovic\/} in October 2008\cite{67:dra}.

Staying with the particle concept problem, there is second interesting issue.
How is it that accumulations of mass arise? It is a convincing idea in cosmology that large accumulations of mass such as is found in a in a galaxy could have arisen by gravitational accretion or by a dispersed distribution of particulate mass falling together as a result of the mutual gravitational attraction of its component particles. Stars could be formed in the same way. This may be what happens in some or many cases but it is possible that such structure are formed and spill out from some unexplained minute but greatly massive singularity. 
Whatever the true situation is, it would seem that a dispersed distribution of negative gravitating particles would not accrete into a single mass because the elements of the distribution would be mutually repulsive. This suggests that negative gravity galaxies of the usual type are unlikely to exist. However, there is another possible twist to this issue and that is how do the mass accumulations within elementary  particles form. Some researchers think that gravity is involved in the structure of elementary particle. They could have been formed by some sort of gravitational collapse and it could be that gravity holds them together. The snag is that gravity is a very weak force and such ideas are not at all convincing. In fact, we have no idea what holds elementary particle together or how they are first formed from {\it energy\/}. In quantum, mechanics however, we do have good ideas about how particle transform between themselves. This does have a bearing on the question of possible galactic size {\it negative\/} mass particles. Dense matter that is found in atoms, molecules, rocks, planets, stars and galaxies is held together by atomic forces of one sort or another. The smaller accumulated parts then held together in the large by positive gravity. This suggests that large accumulations of negative mass particles could be possible if constructed by atomic processes without the intervening spaces found in normal galaxies. I am suggesting that galactic sized molecules could be formed from negatively gravitating particles. Enough speculation, I return to the main issue of the orbiting binary pair and its lessons on fundamentals in the next paragraph.

Another issue of interest is the  question of Newton's {\it Action equals Reaction\/}  principle, his third Law, and how that squares with {\it Action at a distance\/}  in the case of the gravitational interaction between separated positive and negatively characterised  particles. Other than mentioning this issue, I have carried through the work in this paper as though there is no problem. In fact, that is the case, there is no problem this being an important consequence that follows through from Einstein's principle of equivalence in general relativity.
\newpage
To explain this issue, I first quote one popular version of Newton's third law:
\vskip 0.3cm
\centerline{{\it If a first body exerts a force\/} ${\bf f}$ {\it on a second body,\/}}
\centerline{{\it the second body exerts a force\/} -${\bf f}$\  {\it  on the first body.\/}}
 \centerline{${\bf f}$ {\it and\/} -${\bf f}$ {\it are equal in size and opposite in direction.\/}}
\vskip 0.3cm
\noindent Alternatively, in a form nearer to Newton's original presentation:
\vskip 0.3cm
\centerline{{\it All forces occur in pairs, and these two forces are\/}}
\centerline{{\it equal in magnitude and opposite in direction.\/}}
\vskip 0.3cm 
\noindent Newton's original version of this law is not very specific in respect of what a force is and what it is doing but it does seem to be talking about one basic force and its somehow reflected mate such as we all experience when we push an object with our hand and feel the resisting response, presumably the reaction to our action. Such a situation is decidedly local and is essentially all about a single basic force and is very intuitively convincing.  The first definition talks about bodies, says essentially the same as does Newton's law but although it mentions bodies it does not define a body or explain how these two bodies are orientated in space. Of course, fundamental definitions have to be minimal. Neither of these definition obviously apply to bodies separated in space in the way that a binary pair of particle, whatever their gravitational characteristics may be, are separated in space. However, it is usual practice to invoke the action and reaction principle in the standard binary pair case to justify saying the force on one from the other must be of opposite sign to the force from the other to the one. This practice does give a correct result in the usual binary pair case but as we have seen it seems not to work for the mixed mass pair. This situation prompts me to suggest that calling on Newton's third law to support the proposition that the two possible gravitational forces between distant particle from one or the other must be of opposite sign is philosophically incorrect, in spite of it giving a correct result. Before I make this case, let us consider how general relativity theory describes the properties of and incorporates a local gravitational field arising from a distant source. Local {\it acceleration\/} fields generated by distant gravitating particles take precedence over the gravitational forces experienced by any local particle. This is essentially {\it Einstein's Principle of equivalence\/} which manifests itself by the fact that all particles in a given location move with the same acceleration due to local gravity, regardless of their individual masses and, indeed, this is also part of the Newtonian theory structure as described by equations (\ref{w25}) and (\ref{w291}) from which it is clear that the mass of a gravitationally influenced particle cancels out from both side of the equation of motion under gravity. Essentially this means that two spatially separated gravitating particles produce {\it independent\/} acceleration fields in which the other particle moves. Each of the separated particles causes an action at a distance on the other particle but the two particles are acting independently.  This is certainly not reaction of one to the action of the other as implied by Newton's third law from whichever of the two possible directions it may be viewed. However, I have shown above that in the classical case the forces involved do satisfy the condition
\begin{eqnarray}
{\bf f}_1=-{\bf f}_2 , \label{x920}
\end{eqnarray}
which is the condition for action and reaction to be equal in this classical case of a positively characterised pair. Thus if the equal action reaction property holds, how is it reconcilable with my claim that this is not an example of Newton's third law? The answer to this question is that there is a more fundamental simple and kinematic reason of greater generality than Newton's third law. We can see this clearly by examining the equations at (\ref{x13}) and (\ref{x15}) rewritten below in a rearranged form.
\begin{eqnarray}
+ d^2{\bf r}(t)/dt^2&=& \frac{{\bf f}_1}{M_+}=- \frac{{\bf f}_2}{ M_+}\label{x92}\\  
M_+&=&\frac{ m_1^\pm m_2^\pm}{m_1^\pm + m_2^\pm}\label{x92.1}\\
-d^2{\bf r}(t) /dt^2&=& \frac{{\bf f}_1}{M_-}=+ \frac{{\bf f}_2}{M_-}  \label{x94}\\
M_-&=&+\frac{ m_1^\mp m_2^\pm}{m_1^\pm - m_2^\mp},\label{x94.1}
\end{eqnarray}
where $ M_+$ and $ M_-$ are the appropriate {\it reduced\/} centre of mass or centre of gravity masses for the system under consideration. 
\newpage
The relative separation acceleration vectors for a binary system as seen from each particle of the pair for the classical case, the first two, and for the mixed mass case, the second two, can be defined as,
\begin{eqnarray}
{\bf a}_{1+} &=&\frac{{\bf f}_1}{ M_+}\label{x95}\\
{\bf a}_{2+} &=&\frac{{\bf f}_2}{ M_+}\label{x96}\\
{\bf a}_{1-} &=&\frac{{\bf f}_1}{ M_-}\label{x97}\\
{\bf a}_{2-} &=&\frac{{\bf f}_2}{ M_-}.\label{x98}
\end{eqnarray}
If negatively gravitating particles do exist then, according to (\ref{x92}) and (\ref{x94}) all the situations that can occur with a same type gravitating pair or a mixed pair can be summarised as,
\begin{eqnarray}
{\bf a}_{1\pm}= {\scriptstyle \mp}{\bf a}_{2\pm}.\label{x99}
\end{eqnarray}
The formula (\ref{x99}) for relative accelerations has a very simple kinematic and indeed obvious explanation in terms of the geometrical distance between the particle pair, their centre of gravity or their centre of mass. In the case of a {\it same\/} characterised pair. the unit vectors $ \hat {\bf r}_1$ and $ \hat{\bf r}_2$ directed from the members of the pair towards their centre of mass which is between the end points of pairs separation distance and so must be of opposite sign $\hat{\bf r}_1=-\hat {\bf r}_2$. This leads to the ${\bf f}_1 = -{\bf f}_2$ and then implies, with the usual terminology, Newton's action is minus reaction condition. However, in the first case the terminology {\it action equals minus reaction\/} is a misnomer so that the second case when ${\bf f}_1 = {\bf f}_2$ is not in conflict with Newton's third law. 
In the second  case of a {\it differently\/} characterised pair, the unit vectors $\pm \hat {\bf r}_1$ and $\pm \hat{\bf r}_2$ directed from the members of the pair towards their centre of gravity which is outside the pairs separation distance and so  must be of the same sign $\pm\hat{\bf r}_1=\pm\hat {\bf r}_2$. This leads to the ${\bf f}_1 = {\bf f}_2$ and then {\it only seems\/} to imply the {\it violation\/} of Newton's actions is minus reaction condition of the first case. 
\newpage
I suggest that in astronomical physics in the context of separated bodies Newton's law of action and reaction could usefully and more realistically be replaced with the new law:-
\vskip 0.3cm
\centerline{{\it The relative gravitational separation acceleration vectors} $\bf{a}_1$ {\it and} $\bf{a}_2$ }
\centerline{ {\it for two bodies satisfies one or other of the two case,} $\pm$, {\it relation\/}}
\begin{eqnarray}
{\bf a}_{1\pm}={\scriptstyle \mp}{\bf a}_{2\pm}.\label{x100}
\end{eqnarray}
Whether negatively gravitating particles exist or not Newton's law of action and reaction in the astrophysics of separated bodies could usefully be replaced as above with only the $+$ sign in the subcript and the minus sign multiplying.

I have carried through the work in this paper using only Newton's equations of motion but with much guiding influence from general relativity ideas. I think a fully relativistic treatment of this topic would be very complicated and not reveal any substantial deviations from the conclusions that I have arrived at here.
\vskip 0.5cm
\leftline{\bf Acknowledgements}
\vskip 0.5cm
\leftline{I am greatly indebted to Professors Clive Kilmister and} 
\leftline{Wolfgang Rindler for help, encouragement and inspiration}
\leftline{and additionally to Clive and Professor Stephen Barnett for}
\leftline{for drawing my attention to the Bondi negative mass paradox.}

\end{document}